\documentstyle[11pt,epsf]{article}		
\begin{document}
\setcounter{page}{1}
\centerline {\huge\bf Gravitational Lensing in Astronomy}
	\vspace{1.0cm}
	\centerline{\Large Joachim Wambsganss}
	\vspace{0.2cm}
	\centerline{\Large Astrophysikalisches Institut Potsdam}
	\vspace{0.2cm}
	\centerline{\Large An der Sternwarte 16 }
	\vspace{0.2cm}
	\centerline{\Large 14482 Potsdam, Germany}
	\vspace{0.2cm}
	\centerline{\Large\tt jwambsganss@aip.de}
	\vspace{1.5cm}
	\centerline{\Large  Review article for 
			\it Living Reviews in Relativity}
	\vspace{0.2cm}
	\centerline{\Large\tt  http://www.livingreviews.org}

\vspace{2.5cm}

\centerline{\LARGE Abstract}
\vspace{0.5cm}

{\small \large

Deflection of light by gravity was predicted by                    
	General Relativity and observationaly confirmed in 1919. 
In the following decades various aspects of the gravitational lens
	effect were explored theoretically, among them the possibility 
	of multiple or ring-like images of background sources,  
	the use of lensing as a gravitational telescope on very faint
	and distant objects, 
	and the possibility to determine Hubble's constant with
	lensing.
Only relatively recently gravitational lensing became an observational
	science after the discovery of the first doubly imaged quasar 
	in 1979. Today lensing is a booming part of astrophysics.

In addition to multiply-imaged
	quasars, a number of other aspects of lensing 
	have been  discovered since,
	e.g. giant luminous arcs, quasar microlensing, Einstein rings,
	galactic microlensing events, arclets, or 
	weak gravitational lensing. 
	By now literally hundreds of individual
	gravitational lens phenomena are known.

Although still in its childhood, lensing has established itself as a 
	very useful astrophysical tool with some remarkable successes.
It has contributed significant new results in areas as different as 
	the cosmological distance scale,
	the large scale matter distribution in the universe, 
	mass and mass distribution of galaxy clusters, 
	physics of quasars,
	dark matter in galaxy halos, or
	galaxy structure.
Looking at these successes  	
	in the recent past we predict	
	an even more luminous future for gravitational lensing.
}
\vspace{2.5cm}
\begin{center}
1
\end{center}

\pagebreak

\setlength {\topmargin}{-18mm}
%
%
%
%
%
%
%

\section{Introduction}

Within the last 20 years gravitational lensing has changed from 
being considered a geometric curiosity to a helpful and in some ways 
unique tool of modern astrophysics. 
Although the deflection of light at the solar limb was very 
successfully  hailed as the first experiment to confirm 
a prediction of Einstein's
theory of General Relativity in 1919, it took more than half a 
century to establish this phenomenon observationally
in some other environment.
By now almost a dozen different realizations of lensing
are known and observed, and surely more will show up. 

Gravitational lensing -- the attraction of light by matter --
displays a number of attractive features as an academic discipline. 
Its principles are very easy to understand and to explain
due to its being a geometrical effect.
Its ability to produce optical illusions is fascinating
to scientists and laypeople alike. 
And -- most importantly of course -- its usefulness for a number of 
astrophysical problems makes it an attractive tool in many 
branches of astronomy.
All three aspects will be considered below.	

\begin{figure}[hbt]
\unitlength1cm
\begin{picture}(15.0,12.0)   
\put(2.0,-0.5)                    
	{\epsfxsize=12.0cm       
	\epsfbox{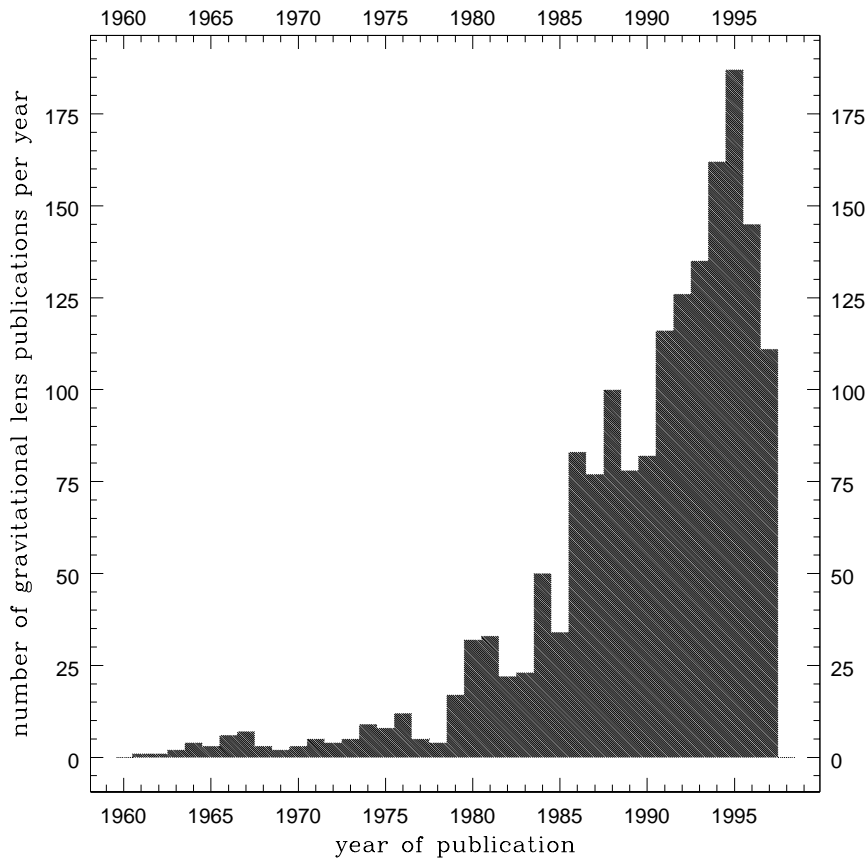}                           
	}   
\end{picture}
\caption{ \label{fig-lens-papers}
	Number of papers on gravitational lensing per year 
	over the last 35 years. 
	This diagram is based on the October 1997 version of the
	lensing bibliography compiled by Pospieszalska-Surdej, Surdej
	and Veron \cite{PSV}.
	The apparent drop after the year 1995 is not real
	but due to incompleteness.}
\end{figure}

In the not quite two decades of its existence as an observational
branch of astrophysics, the field of gravitational lensing has been
continuously growing. 
Every few years a new realisation of the phenomenon was discovered: 
multiple quasars, giant luminous arcs, quasar microlensing, 
Einstein rings,
galactic microlensing, weak lensing, galaxy-galaxy lensing
opened up very different regimes  for the gravitational telescope.
This trend is reflected in a growing number of people working
in the fi2ld.
In Figure \ref{fig-lens-papers} the
number of publications in scientific journals that deal with
gravitational lensing is plotted over time:  
It is obvious that lensing is booming.

%
%

Although there had been a slight sense of disappointment in the
astronomical community a few years ago about the fact that 
lensing had not yet solved all the big problems of astrophysics  right
away (e.g. determination of the Hubble constant; nature of 
dark matter; physics/size of quasars)
this feeling has apparently been reversed by now.
With its many applications and quantitative results lensing 
starts to fulfill its astrophysical promises.

We shall start with a brief look back in time 
and mention some historic aspects of light deflection and lensing
in Chapter \ref{sec-hist}.
We then attempt to explain the basic features 
of gravitational lensing quantitatively, deriving some
of the relevant equations (Chapter \ref{sec-basi}).
A whole variety of lensing observations and phenomena
which curved space-time  provides for us
is presented in Chapter \ref{sec-phen}, e.g.
multiple versions of quasars, 
gigantically distorted  images of galaxies, and 
highly magnified stars.
Along that we explain and  discuss the 
astrophysical applications of lensing which show the use of this tool.
This chapter will be the most detailed one.
Finally in the concluding Chapter \ref{sec-futu} 
we try to extrapolate and speculate about the 
future development of the field.

This article cannot cover the whole field 
of lensing  and its applications. Therefore
we list  here a number of books, proceedings and 
other (partly complementary) review articles on the subject
of gravitational lensing.

The textbook by Schneider, Ehlers \& Falco \cite{SEF}  
contains the most comprehensive presentation of gravitational lensing. 
A new edition is underway.
The book by Bliokh \& Minakov \cite{BM89} 
on gravitational lensing
is still only available in Russian.
A new book currently in press  by Petters, Levine \& Wambsganss 
\cite{PLW} 
treats mainly the mathematical aspects of lensing, in particular
its applications to singularity theory.

The contributions to the most important conferences
on gravitational lensing in the last few years were all
published: 
Swings \cite{liege83} edited the Proceedings on the first 
		conference on lensing in Li\`ege in 1983;
Moran   et al. \cite{MIT} are the editors of the MIT
		workshop on lensing in 1988;
Mellier et al. \cite{toulouse} of the Toulouse conference in 1989;
Kayser et al. \cite{hamburg} of the Hamburg meeting in 1991;
Surdej et al. \cite{liege93} of the Li\`ege conference in 1993;
Kochanek \& Hewitt \cite{melbourne}  of  the IAU Symposium 173 in 
	Melbourne in 1995;
Jackson \cite{Jac97} of the Jodrell Bank Meeting 
	``Golden Lenses" in 1997.

There exist a number of excellent reviews on gravitational lensing.
Blandford \& Kochanek \cite{BK87} give a nice introduction on
the theory of lensing. The optical aspects of lensing
are derived elegantly in \cite{Bla89}.
The presentation of Blandford \& Narayan \cite{BN92} emphasizes
in particular
the cosmological applications of gravitational lensing.
The review by Refsdal \& Surdej \cite{RS94} contains a section on
optical model lenses that simulate the lensing action
of certain astrophysical objects.
A recent review article by Narayan \& Bartelmann \cite{NB96} summarizes
in a very nice and easy-to-understand way the basics and the latest
in the gravitational lens business.
In the sections below some more specific review articles will be 
mentioned.

%
%
%
%
%
%
%
%
\section{History of Gravitational Lensing}
\label{sec-hist}

The first written account of the deflection of light by gravity  
appeared in the ``Berliner Astronomisches Jahrbuch auf das Jahr 1804"
in an article entitled: 
``Ueber die Ablenkung eines Lichtstrals von seiner geradlinigen
Bewegung, durch die Attraktion eines Weltk\"orpers, an welchem er nahe
vorbeigeht"\footnote{
``On the Deflection of a Light Ray from its Straight Motion
due to the Attraction of a World Body  which it Passes Closely"}
\cite{Sol1801}.
Johann Soldner 
 -- a German geodesist, mathematician and astronomer, then working
 at the Berlin Observatory -- 
 explored  this effect and inferred that a 
light ray close to the solar limb would be deflected by 
an angle $\tilde\alpha = 0.84$ arcsec. 
It is very interesting to read how carefully and cautiously he 
investigated this idea and its consequences on
practical astronomy.

In the year 1911 -- more than a century later -- 
Albert Einstein  \cite{Ein11} 
directly addressed the influence of gravity on light
(``\"Uber den Einflu{\ss} der
Schwerkraft auf die Ausbreitung des Lichtes"\footnote{``On the
Influence of Gravity on the Propagation of Light"}).
At this time the General Theory of Relativity was not
fully developed yet. 
This is the reason why Einstein obtained
-- unaware of the earlier result -- the same value
for the deflection angle
as Soldner had calculated with Newtonian physics.
In this paper Einstein found   
$\tilde\alpha = 2 G M_{\odot} / c^2 R_{\odot}  =  0.83$ arcsec
for the deflection angle of a ray grazing the sun
(here $M_{\odot}$ and $R_{\odot}$ are the mass and the 
radius of the sun,
$c$ and $G$ are the velocity of light and the gravitational constant,
respectively).
Einstein emphasized his wish 
that astronomers investigate this question
 (``Es w\"are dringend zu
w\"unschen, da{\ss} sich Astronomen der hier aufgerollten Frage
ann\"ahmen, auch wenn die im vorigen gegebenen \"Uberlegungen
ungen\"ugend fundiert  oder gar abenteuerlich erscheinen
sollten."\footnote{``It would be very desirable that astronomers 
	address the question unrolled here, even if the considerations
	should seem to be insufficiently founded or entirely 
	speculative."}).
Recently it was discovered that Einstein had derived the lens equation,
the possibility of a double image and the magnifications of the images
in a notebook in the year 1912 \cite{Ren97}.
In 1913 Einstein even contacted directly the director of the 
Mt. Wilson Observatory -- George Ellery Hale -- and 
asked him whether it would be possible to measure positions of 
stars near the sun during the day in order to establish
the deflection effect of the sun (see \cite{AIP}).
	Einstein's hand-drawn sketch marks the  
	the ``wrong" value of 0.84 arcsec. 
	%
	%

There actually were plans to test Einstein's wrong prediction
of the deflection angle 
during a solar eclipse in 1914 on the Russian Crimea
peninsula. 
However, when the observers were already in Russia,
World War I broke out and they were captured by Russian soldiers
\cite{Bre82}.
So -- fortunately for Einstein -- the measurement of the
deflection angle at the solar limb had to be postponed 
for a few years.

With the completion of the General Theory of Relativity
Einstein was the first to derive the correct
deflection angle $\tilde \alpha$ of a light ray
passing at a distance $r$ from an object of mass $M$ as
\begin{equation}
	\tilde \alpha  = {4 G M \over c^2}{1 \over  r}.
\end{equation}
The additional factor of two (compared to the ``Newtonian" value) reflects
the spatial curvature which is missed if photons are just treated as
particles. 
With the solar values  for radius and mass Einstein obtained
\cite{Ein16,Ein22}: 	
\begin{equation}
      \tilde \alpha_{\odot} = { {4 G M_{\odot}} \over  { c^2}} 
			{1 \over  R_{\odot}} =  1.74 {\rm \ arcsec}.
\end{equation}
It is common wisdom now that the determination 
of this value to within 20\% 
during the solar eclipse in 1919 by Arthur Eddington and his group
was the second observational confirmation
of General Relativity \cite{Dys20}     
and the basis of Einstein's huge popularity
starting in the 1920s (the first one had been the
explanation of Mercury's perihelion shift).
Recently the value predicted by
Einstein  was confirmed to an accuracy 
better than 0.02 \% \cite{Leb95}.

In the following decades light deflection or gravitational
lensing was only very rarely the topic of a research paper:
In 1924 Chwolson \cite{Chw24} 
mentioned the idea of a ``fictitous double star" and the mirror-reversed
nature of the secondary image. 
He also mentioned the symmetric case of star exactly behind star,
resulting in a circular image.
Einstein also reported in 1936 about the appearance of 
a ``luminous circle" for perfect alignment between source
and lens \cite{Ein36}, 
and of two magnified images for slightly displaced
positions\footnote{As stated above,
	only very recently  it was shown that Einstein
	had derived these equations  as early as 1912, 
	but did not bother to publish them \cite{Ren97}.}.
Today such a lens configuration  is called ``Einstein-ring", 
although more correctly it should be called ``Chwolson-ring".
Influenced by Einstein, Fritz Zwicky \cite{Zwi37a,Zwi37b} 
pointed out in 1937 that galaxies
(``extragalactic nebulae") are much more likely to be gravitationally
lensed than stars and that one can use the gravitational
lens effect as a ``natural telescope".

In the 1960s a few partly independent theoretical studies showed the 
usefulness of lensing for 
astronomy \cite{Kli63,Lie64,Lin67,Met63,Ref64a,Ref64b}.  
In particular Sjur Refsdal 
derived the basic equations of gravitational 
lens theory and subsequently
showed how the gravitational lens effect
can be used to determine Hubble's constant
by measuring the time delay between two lensed images.
He followed up this work with interesting applications of lensing
\cite{Ref66a,Ref66b,Ref70}.
The mathematical foundation of how a light bundle is distorted
on its passage throught the universe
had been derived in the context of gravitational radiation
even before \cite{Sac61}.

Originally gravitational lensing was discussed for stars or for
galaxies. When in the 1960s quasars were discovered, 
Barnothy \cite{Bar65}
was the first to connect them with 
the gravitational lens effect. 
In the late 60s/early 70s a few groups and individuals
explored various aspects of lensing further,
e.g.: statistical effects of local inhomogeneities
on the propagation of light \cite{Gun67a,Gun67b,PG73};
lensing applied to quasars and clusters of galaxies 
\cite{CK75,Noo71,San71};
development of a formalism for transparent lenses \cite{BK75,Cla72};
or the effect of an inhomogeneous universe
on the distance-redshift relations \cite{DR73}.

But only in the year 1979 the whole field 
received a real boost when the first double
quasar was discovered and confirmed to be a real gravitational lens
by Walsh, Carswell \& Weymann \cite{WCW}.  
This discovery and the development of lensing since then
will be described in Chapter \ref{sec-phen}.

There are a few historic accounts of lensing which are more
detailed than the one presented here, e.g.  
in \cite{Rob83},
in \cite{SEF}, or
in \cite{Wam90}.
The complete history of gravitational lensing has to be written yet.

%
%
%
%
%
%
%
\section{Basics of Gravitational Lensing}
\label{sec-basi}

The path,  the size and the cross section
of a light bundle propagating through spacetime 
in principle are affected by all the matter between the light
source and the observer. 
For most practical purposes  we can assume  that
the lensing action is dominated by a single matter inhomogeneity
at some location between source and observer. 
This is usually called the ``thin lens approximation": 
all the action of deflection is thought  to take place at a 
single distance. 
This approach is valid only if the relative velocities of lens, source
and observer are small compared to the velocity of light 
$v \ll c$ and if the Newtonian potential is small $|\Phi| \ll c^2$.
These two assumptions are
justified in all astronomical cases of interest. 
The size of a galaxy, e.g.,  is of order 50 kpc, even a cluster of 
galaxies is not much larger than 1 Mpc. This ``lens thickness" is
small compared to the typical distances of order few Gpc
between observer and lens or lens and background quasar/galaxy, 
respectively.  We  assume  that the underlying spacetime is
well described by a perturbed Friedmann-Robertson-Walker 
metric\footnote{A detailed description of optics in curved spacetimes 
	and a derivation of the lens equation from Einstein's field 
	equations can be found in Chapters 3 and 4 of \cite{SEF}.}:


\subsection*{Lens Equation }

\begin{figure}[bth]
	\unitlength1cm
	\begin{picture}(14.0,11.0)   
	\put(0.5,-0.5)               
	{\epsfxsize=12.0cm           
		\epsfbox{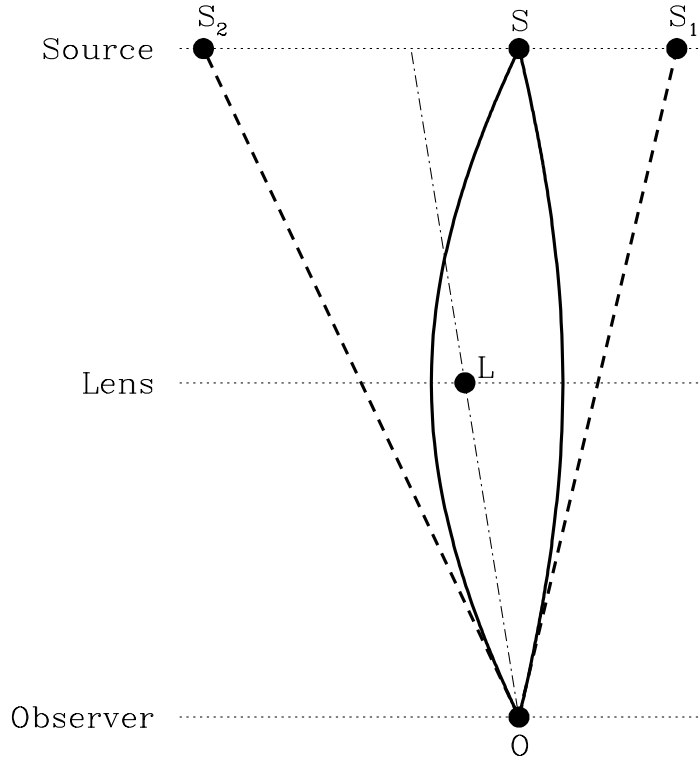}                           
			}   
	\end{picture}
\caption{\label{fig-setup}
	Setup of a gravitational lens situation: The lens $L$ located 
	between source $S$ and observer $O$ produces two images $S_1$ 
	and $S_2$ of the background source.  }
\end{figure}

The basic setup for such a simplified 
gravitational lens scenario involving a point source
and a point lens is displayed in
Figure \ref{fig-setup}. 
The three ingredients in such a lensing situation  are the
source S, the lens L, and the observer O.
Light rays emitted from the source are deflected by the lens.
For a point-like lens, there will always be (at least) two images
S$_1$ and S$_2$  of the source.  With external shear -- due to
the tidal field of objects outside but near the light bundles -- 
there can be more images.
The observer sees the images in directions corresponding to the 
tangents to the real incoming light paths.

%
%

\begin{figure}[bth]
	\unitlength1cm
	\begin{picture}(14.0,11.0)   
	\put(2.5,-0.9)               
	{\epsfxsize=12.0cm           
		\epsfbox{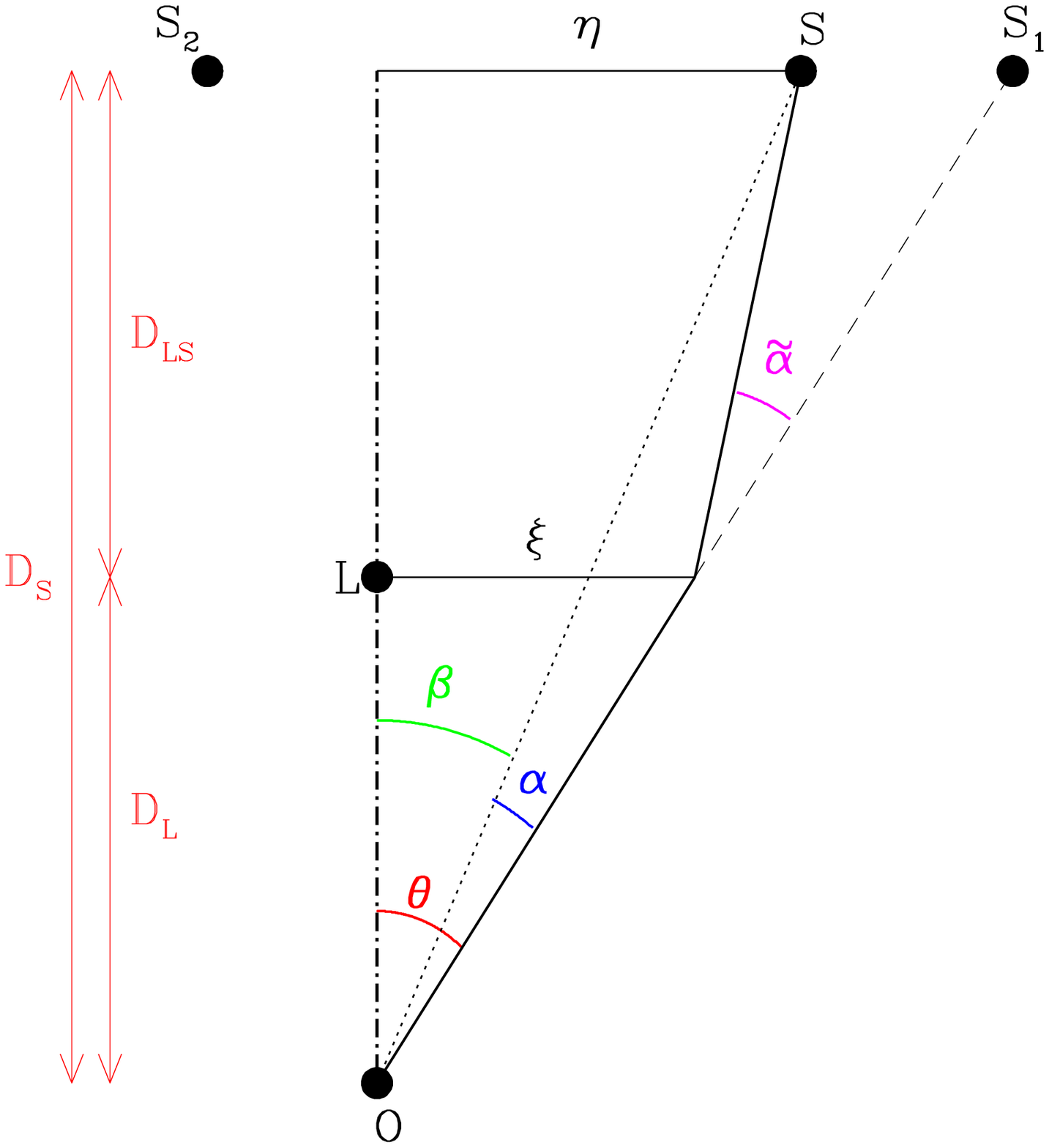}                           
			}   
	\end{picture}
\caption{\label{fig-setup-eq}
	The relation between the various angles and distances involved
	in the lensing setup can be derived for the case 
	$\tilde \alpha \ll 1$
	and formulated in the lens equation (\ref{eq-lenseq}).
}
\end{figure}

In Figure \ref{fig-setup-eq} the corresponding angles and angular
diameter distances $D_L$, $D_S$, $D_{LS}$ are 
indicated\footnote{In cosmology the various methods to define
	distance diverge (see, e.g. Chapter 14.4 of \cite{Wei72})
	or 3.5 of \cite{SEF}). 
	The relevant distances for gravitational lensing 
	are the angular diameter distances, see \cite{NB96}.}. 
In                  the thin-lens approximation
the hyperbolic paths are approximated by their asymptotes.
In the circular-symmetric case the deflection angle is 
given as

\begin{equation}
\label{eq-angle}
\tilde \alpha (\xi)  =  { { 4 G M(\xi)} \over {c^2} } { 1  \over \xi }.
\end{equation}
where $M(\xi)$ is the mass inside a radius $\xi$. In this depiction
the origin is chosen at the observer. 
From the diagram it can be seen that the following relation holds:
\begin{equation}
                \theta D_S = \beta D_S + \tilde \alpha  D_{LS} 
\end{equation}
(for $\theta$,  $\beta$, $\tilde \alpha \ll 1$;  this condition
is fulfilled 
in practically all astrophysically relevant situations).
With the definition of the reduced deflection angle as 
$\alpha (\theta)  =   ( D_{LS} / D_{S} ) \tilde \alpha (\theta)$,
this can be expressed as:
\begin{equation}
	\label{eq-lenseq}
       \beta = \theta - \alpha (\theta).
\end{equation}
This relation between the positions of images and source 
can easily be derived for a non-symmetric mass distribution as well. 
In that case all angles are vector-valued. 
The two-dimensional {\bf lens equation}   then reads:
\begin{equation}
	\label{eq-lenseq-vec}
       \vec\beta = \vec\theta - \vec\alpha (\vec\theta).
\end{equation}
%
%

%
%


\subsection*{Einstein Radius}

For  a point lens of mass $M$ the deflection angle is given
by equation (\ref{eq-angle}). Plugging into equation (\ref{eq-lenseq})
and using the relation
$\xi = D_L \theta $ (cf. Figure \ref{fig-setup-eq}) one obtains:
\begin{equation}
       \beta(\theta)  = 
       \theta - { D_{LS} \over D_L D_S} { 4 G M \over c^2 \theta}.
\end{equation}
For the special case in which the source lies exactly behind the
lens ($\beta = 0$), due to the symmetry a ring-like image occurs
whose angular radius is called {\bf Einstein radius} $\theta_E$:

\begin{equation}
       \theta_E  = \sqrt{ 				
			{ 4 G M \over c^2}  
			{D_{LS} \over D_L D_S}.
				}      
\end{equation}
The Einstein radius defines the angular scale for a lens
situation. For a massive galaxy with a mass of
$M = 10^{12} M_{\odot}$ at a redshift of $z_L = 0.5$
and a source at redshift $z_S = 2.0$  
(we used here $H = 50 $km sec$^{-1}$ Mpc$^{-1}$ as  the value
of the Hubble constant and an Einstein-deSitter universe)
the Einstein radius is
\begin{equation}
	\label{eq-angle-gal}
       \theta_E  \approx 1.8  \ \sqrt{ M \over 10^{12} M_\odot } \ {\rm arcsec}
\end{equation}
(note that for cosmological distances in general 
$D_{LS} \ne D_S - D_L$!).
For a galactic microlensing scenario in 
which stars in the disk of the Milky Way act as lenses 
for bulge stars close to the center of the Milky Way, 
the scale defined by the Einstein radius is
\begin{equation}
       \theta_E  \approx 0.5  \ \sqrt{ M \over M_{\odot} }   \ {\rm milliarcsec}.
\end{equation}
An application and some illustrations of the point lens case can
be found in Section \ref{sec-gal-micro} on galactic microlensing.


\subsection*{Critical Surface Mass Density }

In the more general case of 
a three-dimensional mass distribution of an extended
lens, the density $\rho(\vec r)$ can be projected along the line
of sight onto the lens plane to 
obtain the two-dimensional
surface mass density distribution $\Sigma(\vec\xi)$, as 
\begin{equation}
\Sigma(\vec\xi) = \int_0^{D_S} \rho(\vec r) dz.
\end{equation}
Here $ \vec r $  is a three-dimensional vector in space, and
     $ \vec \xi$  is a two-dimensional vector in the lens plane.
The two-dimensional deflection angle $\vec{\tilde \alpha}$
is then given as the sum over all mass elements
in the lens plane:
\begin{equation}
\vec{ \tilde \alpha} (\vec{\xi})=  { {4 G} \over {c^2}}
	\int { { (\vec\xi - \vec\xi') \Sigma(\vec\xi')}  \over 
	{ |\vec\xi - \vec\xi'|^2 }  }d^2\xi'.
\end{equation}
For a finite circle with constant surface mass 
density $\Sigma$ the deflection angle can be written:
\begin{equation}
    {        \alpha (\xi) } =   { D_{LS} \over D_S} 
			{ 4 G \over c^2    } 
			{ \Sigma \pi \xi^2 \over     \xi} 
\end{equation}
With $\xi = D_L \theta$ this simplifies to
\begin{equation}
    {        \alpha (\theta) } =   { 4 \pi G \Sigma \over c^2 } 
	{ D_L D_{LS} \over D_S} \theta.
\end{equation}
With the definition of the {\bf critical surface mass density}
$\Sigma_{\rm crit}$ as 
\begin{equation}
\label{eq-crit}
\Sigma_{\rm crit}  =  { c^2  \over {4 \pi G }} { D_S  \over {D_L D_{LS} }}
\end{equation}
the deflection angle for a such a mass distribution can be expressed as
\begin{equation}
    { \tilde \alpha (\theta) } = { \Sigma \over \Sigma_{\rm crit} } \theta.
\end{equation}
The critical surface mass density is given by the lens mass $M$
``smeared out" over the area of the Einstein ring:  
$\Sigma_{crit} = M / (R_E^2 \pi)$, where  $R_E = \theta_E D_L$.
The           value of the  critical surface mass density
is roughly $\Sigma_{\rm crit}  \approx 0.8$ g cm$^{-2}$ 
for lens and source redshifts of $z_L = 0.5$ and
$z_S = 2.0$, respectively.
For an arbitrary mass distribution  the condition
$ \Sigma > \Sigma_{\rm crit}$ at any point is sufficient to
produce multiple images.


\subsection*{Image Positions and Magnifications }

The lens equation (\ref{eq-lenseq}) can be re-formulated
in the case of a single point lens:

\begin{equation}
	\label{eq-lenseq-point}
       \beta = \theta - { \theta_E^2  \over \theta}.
\end{equation}
Solving this for the image positions $\theta$ one finds that
an isolated point source always produces two images of a background
source.  The positions of the images are given by the
two solutions:
\begin{equation}
	\label{eq-lenseq-images}
       \theta_{1,2} = {1 \over 2 } 
       \left( \beta  \pm \sqrt{\beta^2 + 4 \theta_E^2} \right).
\end{equation}
The magnification of an image is defined by the ratio between the
solid angles of the image and the source, since the surface
brightness is conserved. 
Hence the magnification $\mu$ is given as
\begin{equation}
	\label{eq-magn}
       \mu  = {\theta \over \beta} {d \theta \over d \beta}.
\end{equation}
In the symmetric case above the image magnification
can be written as (by using the lens equation):
\begin{equation}
       \mu_{1,2}  
       = 
       \left( 1 - \left[ {\theta_E \over \theta_{1,2}} \right]^4  \right)^{-1}
       = 
           { u^2 + 2 \over 2 u \sqrt{u^2 + 4 } } \pm {1 \over 2}
\end{equation}
Here we defined $u$ as the ``impact parameter", the angular
separation between lens and source in units of the Einstein
radius: $u = \beta / \theta_E$. 
The magnification of one image (the one inside
the Einstein radius) is negative. This  means it has
negative parity:  it is mirror-inverted.
For $\beta \rightarrow 0$ the magnification diverges: in the
limit of geometrical optics the Einstein ring of a point source
has infinite magnification\footnote{Due to the fact
	that physical objects have a finite size, and also
	because at some limit wave optics has to be applied,
	in reality the magnification stays finite.}!
The sum of the absolute values of the two image magnifications
is the measurable total magnification $\mu$:
\begin{equation}
	\label{eq-magn-tot}
       \mu = |\mu_{1}| + |\mu_2| = { u^2 + 2 \over   u \sqrt{u^2 + 4 } }.
\end{equation}
Note that this value is (always) larger than one\footnote{This
	does not violate energy conservation, since this is
	the magnification relative to an ``empty" universe and
	not relative to a ``smoothed out" universe. This issue
	is treated in detail in, e.g., \cite{Sch84} or in 
	Chapter 4.5 of \cite{SEF}.}!
The difference between the two image magnifications is  unity:
\begin{equation}
       \mu_{1} + \mu_2 = 1.
\end{equation}
%
%


\subsection*{(Non-)Singular Isothermal Sphere }
A handy and popular model for galaxy lenses is the singular isothermal
sphere with a three-dimensional density distribution of 
\begin{equation}
       \rho(r) = { \sigma_v^2 \over 2 \pi G} {1 \over r^2},
\end{equation}
where        $\sigma_v$ is the one-dimensional velocity dispersion.
Projecting the matter on a plane one obtains the circularly-symmetric
surface mass distribution
\begin{equation}
       \Sigma(\xi) = { \sigma_v^2 \over 2 G} {1 \over \xi }.
\end{equation}
With \ $M(\xi) = \int_0^\xi \Sigma(\xi') 2 \pi \xi' d\xi' $ \ 
plugged into equation (\ref{eq-angle}) 
one obtains the  deflection
angle for an isothermal sphere, which is a constant (i.e. independent
of the impact parameter $\xi$):
\begin{equation}
       \tilde \alpha(\xi) = 4 \pi { \sigma_v^2 \over c^2 }.
\end{equation}
In ``practical units"  for the
velocity dispersion this can be expressed as:
\begin{equation}
       \tilde \alpha(\xi) 
	= 1.15 \left( \sigma_v \over 200 {\rm km s}^{-1} \right)^2  
	{\rm arcsec} .
\end{equation}

Two generalizations of this isothermal model are commonly used: 
Models with finite cores 
are more realistic for (spiral) galaxies. 
In this case the deflection angle is modified
to (core radius $\xi_c$):
\begin{equation}
       \tilde \alpha(\xi) 
= 4 \pi { \sigma_v^2 \over c^2 } {\xi \over {(\xi_c^2 + \xi^2)^{1/2}}}. 
\end{equation}
Furthermore, a realistic galaxy lens usually is not perfectly
symmetric but is slightly elliptical. Depending on
whether one wants an elliptical mass distribution or
an elliptical potential, various formalisms have been suggested.
Detailed treatments of elliptical
lenses can be found in \cite{bar98,BK87,KK93,KK98,Kor94,Schramm}.


\subsection*{Lens Mapping }

In the vicinity of an arbitrary point, 
the lens mapping as shown in equation (\ref{eq-lenseq-vec})
can be described by its Jacobian matrix $\cal A$:
\begin{equation}
       {\cal A} = { \partial \vec \beta \over \partial \vec \theta} = 
         \left( \delta_{ij} 
		- {\partial \alpha_i (\vec \theta) \over \partial \theta_j} \right)
	 = 
         \left( \delta_{ij} 
	 - {\partial^2 \psi (\vec \theta) 
		\over \partial \theta_i \partial \theta_j} \right).
\end{equation}
Here we made use of the fact (see \cite{Bla86,Sch85}),
that the deflection angle can be expressed
as the gradient of an effective two-dimensional scalar
potential $\psi$: \ $\vec \nabla_{\theta} \psi = \vec \alpha$,
where 
\begin{equation}
	\psi(\vec \theta) = { D_{LS} \over {D_L D_S} } {2 \over c^2}
	\int \Phi (\vec r) dz
\end{equation}
and $\Phi (\vec r) $ is the Newtonian potential of the lens.

\begin{figure}[tbm]
	\unitlength1cm
	\begin{picture}(15.0,8.0)   
	\put(-0.5,-0.5)               
	{\epsfxsize= 8.0cm           
		\epsfbox{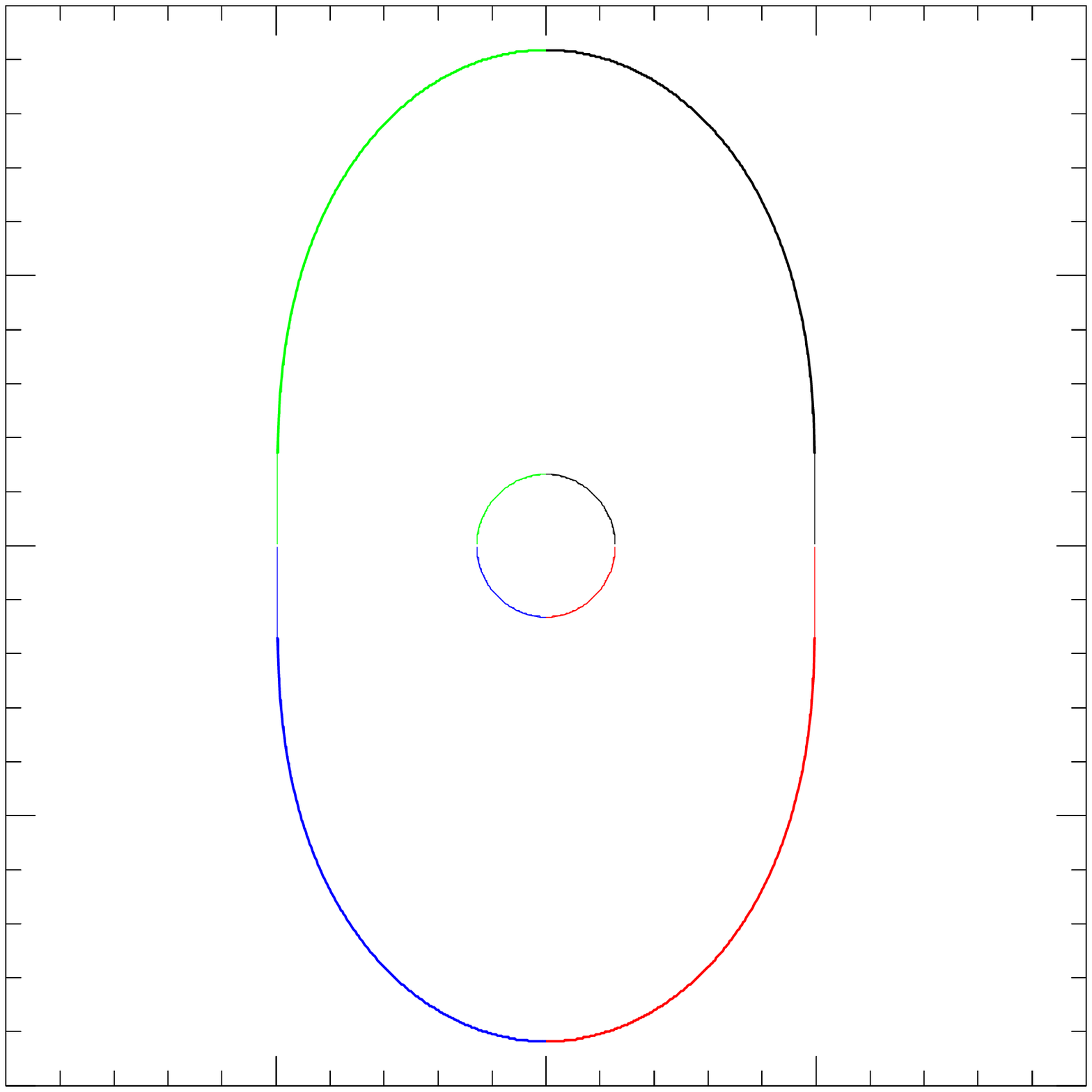}                          
			}   
	\put(7.5,-0.5)               
	{\epsfxsize= 8.0cm           
		\epsfbox{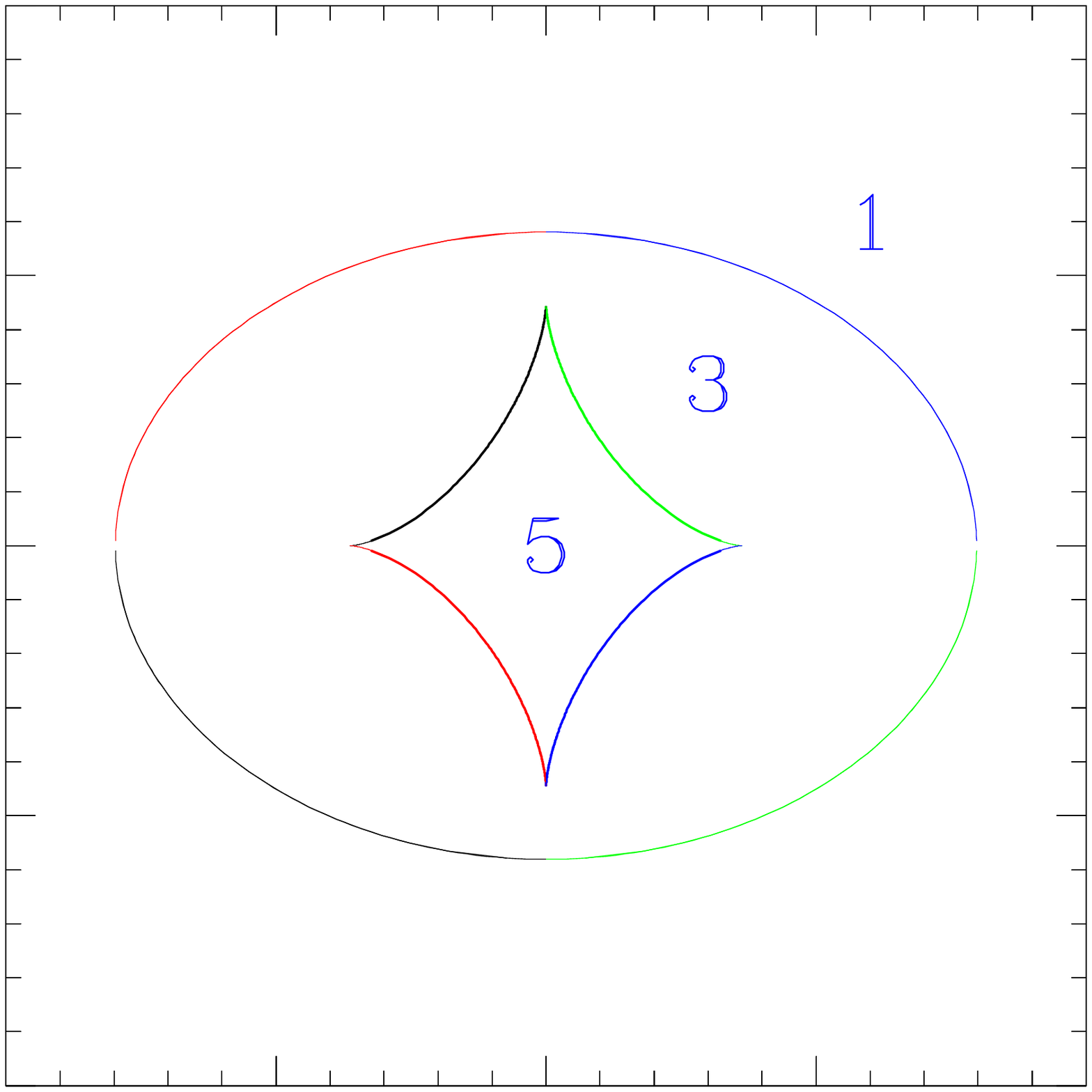}                          
			}   
	\end{picture}
\caption{\label{fig-lens-elliptical}
	The critical curves (left) and caustics (right) for an 
	elliptical lens.  The numbers in the right panels identify
	regions in the source plane that
	correspond to 1, 3 or 5 images, respectively. The
	smooth lines in the right hand panel are called
	fold caustics; the tips at which in the
	inner curve two fold caustics
	connect are called cusp caustics.
}
\end{figure}

The determinant of the Jacobian $\cal A$  is the inverse of the 
magnification:
\begin{equation}
	\label{eq-magn-det}
	\mu = {1 \over \det {\cal A}}.
\end{equation}
Let us define
\begin{equation}
 \psi_{ij} = {\partial^2 \psi  \over \partial \theta_i \partial \theta_j}.
\end{equation}
The Laplacian of the effective potential
$\psi$ is twice the convergence:
\begin{equation}
	\label{eq-kappa}
 \psi_{11} + \psi_{22}  = 2 \kappa = {\rm tr } \    \psi_{ij} .
\end{equation}
With the definitions of the components of the external shear $\gamma$:
\begin{equation}
	\label{eq-gamma1}
  \gamma_1 (\vec \theta) = {1 \over 2} (\psi_{11} - \psi_{22}) = 
  \gamma (\vec \theta) \cos [2 \varphi (\vec \theta)] 
\end{equation}
and 
\begin{equation}
	\label{eq-gamma2}
  \gamma_2 (\vec \theta) =  \psi_{12}   =  \psi_{21} = 
  \gamma (\vec \theta) \sin [2 \varphi (\vec \theta)] 
\end{equation}
(where the angle $\varphi$ reflects
the direction of the shear-inducing
tidal force relative
to the coordinate system)
the Jacobian matrix can be written 
\begin{equation}
  {\cal A} = 
 	 \left( 
	\begin{array}{cc} 
			1 - \kappa - \gamma_1 & -\gamma_2 \\
                           -\gamma_2  &  1 - \kappa + \gamma_1 \\
	\end{array} 
	\right)
	=  ( 1 - \kappa )
 	 \left( 
	\begin{array}{cc} 
			1 & 0 \\
			0 & 1 \\ 
	\end{array} 
	\right)
	- \gamma
 	 \left( 
	\begin{array}{lr} 
			\cos 2\varphi  &  \sin 2\varphi  \\
			\sin 2\varphi  & -\cos 2\varphi  \\ 
	\end{array} 
	\right).
\end{equation}
The magnification can now be expressed as a function of the local
convergence $\kappa$ and the local shear $\gamma$:
\begin{equation}
\mu = ( \det {\cal A})^{-1}  =  { 1 \over (1-\kappa)^2 - \gamma^2}.
\end{equation}
Locations at which \ \ $\det A = 0$ \ \  have formally infinite
magnification. They are called {\bf critical curves} in the lens
plane. The corresponding locations in the source plane are the
{\bf caustics}.  For spherically symmetric mass distributions,
the critical curves are circles. For a point lens,  the caustic
degenerates into a point. 
For elliptical lenses or spherically
symmetric lenses plus external shear, 
the caustics can consist of cusps and folds.
In Figure \ref{fig-lens-elliptical}
the caustics and critical curves for an elliptical
lens with a finite core are displayed.

%
%


\subsection*{Time delay and ``Fermat's" Theorem}

The deflection angle is 
the gradient of an  effective lensing
potential $\psi$ 
(as was first shown by \cite{Sch85}; see also \cite{Bla86}).
Hence the lens equation can be rewritten as
\begin{equation}
(\vec \theta - \vec \beta) - \vec \nabla_\theta \psi  = 0
\end{equation}
or
\begin{equation}
\label{eq-nabla}
\vec \nabla_{\theta} 
	\left( {1 \over 2} (\vec \theta - \vec \beta)^2  - \psi \right) = 0.
\end{equation}
The term in brackets appears as well in the physical 
time delay function for gravitationally lensed images:
\begin{equation}
\label{eq-time-delay}
\tau(\vec \theta, \vec \beta) =  \tau_{\rm geom} + \tau_{\rm grav} =
	{ 1 + z_L \over c } { D_L D_S \over D_{LS}}
  \left( {1 \over 2} (\vec \theta - \vec \beta)^2  - \psi(\theta)  \right).
\end{equation}
This time delay surface is a function of 
the image geometry ($\vec \theta$, $\vec \beta$), 
the gravitational potential $\psi$,
and the distances $D_L$, $D_S$, and $D_{LS}$.
The first part 
-- the geometrical time delay $\tau_{\rm geom}$ --
reflects the extra path length compared to the direct
line between observer and source. 
The second part 
-- the gravitational time delay $\tau_{\rm grav}$ --
is the retardation due to gravitational potential of 
the lensing mass (known and confirmed
as Shapiro delay in the solar system).
From equations  (\ref{eq-nabla}), (\ref{eq-time-delay}) 
it follows that the 
gravitationally  lensed  images appear at locations that correspond
to extrema in the light travel time, which reflects 
Fermat's principle in gravitational-lensing optics.

The (angular-diameter) distances that appear
in equation (\ref{eq-time-delay}) depend on the value
of the Hubble constant \cite{Wei72}; therefore
it is possible to determine
the latter by measuring the time delay between different
images and using a good model for the effective
gravitational potential $\psi$ of the lens (see
\cite{Kun97,Ref64b,WS97} and Section \ref{sec-mult}).

%
%

This chapter followed heavily the elegant presentation of 
the basics of lensing in Narayan \& Bartelmann \cite{NB96}.
Many more details can be found there. 
More complete derivations of the lensing properties
are also provided in all the introductory
texts mentioned in Chapter 1, in particular in \cite{SEF}.
More on the formulation of gravitational lens theory in terms
of time-delay and Fermat's principle can be found in
Blandford \& Narayan \cite{Bla86}    and
Schneider \cite{Sch85}.
Discussions of the concept of ``distance" in relation to
cosmology/curved space can be found in chapter 3.5 of \cite{SEF}
or chapter 14.4 of \cite{Wei72}.

%
%
%
%
%
%
%
%
%
\section{Lensing Phenomena}
\label{sec-phen}

In this chapter we describe different groups
of gravitational lens observations. 
The subdivision is pragmatic rather than entirely logical.
It is done partly by lensed object,
or by lensing object, or by lensing strength. 
The ordering roughly reflects
the chronological appearance of
different sub-disciplines to lensing.
The following sections are on
\begin{itemize}
\item Multiply-imaged quasars
\item Quasar microlensing  
\item Einstein rings 
\item Giant luminous arcs and arclets 
\item Weak lensing / Statistical lensing
\item Cosmological aspects of (strong) lensing 
\item Galactic microlensing
\end{itemize}
Comprehensive reviews  could be
written on each separate subject listed here. Hence the treatment
here can be only very cursory.


\subsection{Multiply-imaged quasars}
\label{sec-mult}

In the year 1979 gravitational lensing became an
observational science when the double quasar Q0957+561 was 
discovered - the first example of a lensed object \cite{WCW}.
The discovery itself happened rather by accident; 
the discoverer Dennis Walsh describes in a nice account how 
this branch of astrophysics came into being \cite{Wal88}. 

It was  not entirely clear at the beginning, though, whether the
two quasar images  really were 
an illusion provided by curved space-time -- or rather
physical twins.
But intensive observations soon confirmed the almost identical
spectra. The intervening ``lensing" galaxy was found, and the
``supporting" cluster was identified as well. 
Later very similar lightcurves of the two images (modulo offsets
in time and magnitude) confirmed this system beyond any doubt as 
a bona fide gravitational lens.

By now about two dozen multiply-imaged quasar systems
have been found, plus another ten good candidates
(updated tables of multiply-imaged
quasars and gravitational lens candidates 
are provided, e.g., by the CASTLE group \cite{CASTLE}).
This is not really an exceedingly big number, considering
a 20 year effort to find lensed quasars. 
The reasons for this ``modest" success rate is 
\begin{enumerate}
\item  Quasars are rare and not easy to find 
	(by now roughly $10^4$ are  known).
\item The fraction of quasars that is lensed is 
	small (less than one percent).
\item It is not trivial at all to {\it identify}  the lensed (i.e. 
	multiply-imaged) quasars among the known ones.
\end{enumerate}
Gravitationally lensed quasars come in a variety of classes:
double, triple and quadruple systems;
symmetric and asymmetric image configurations are known.

A recurring problem connected with double quasars is the 
question whether they are two images of a single source or 
rather a physical association of two objects (with three or
more images it is more and more likely that it is lensed system).
Few systems are as well established as the double quasar Q0957+561;
but many are considered ``safe" lenses as well. Criteria
for ``fair", ``good", or ``excellent" lensed quasar candidates 
comprise the following:

\begin{itemize}
\item There are two or more point-like images of very similar optical color.
\item Redshifts  (or distances) of both quasar images are identical 
	or {\it very} similar.
\item Spectra of the various images are  identical 
	or {\it very} similar to each other.
\item There is a lens (most likely a galaxy) found between the images,
	with a measured redshift much smaller than the quasar redshift
	(for a textbook example see Figure \ref{fig-lens-galaxy}).
\item  If the quasar is intrinsically variable, the fluxes measured 
	from the two (or more) images follow a 
	very similar light curve, except for certain lags -- 
	the time delays -- and an overall
	offset in brightness (cf. Figure \ref{fig-0957-light}).
\end{itemize}
For most of the known multiple quasar systems only some of the
above criteria are fully confirmed. 
And there are also good reasons not to require perfect agreement 
with this list:
e.g., the lensing galaxy could be superposed to one quasar
image and make the quasar appear extended; 
color/spectra could be affected by dust absorption in the
lensing galaxy and appear not identical; the lens could be too faint
to be detectable (or even a real dark lens?); 
the quasar could be variable on time scales shorter than
the time delay; 
microlensing can affect the lightcurves of the images differently.  
Hence it is not easy to say how many 
gravitationally lensed quasar systems exist.  The answer depends
on the amount of certainty one requires.
In a recent compilation Keeton \& Kochanek \cite{KK96} put
together 29 quasars as lenses or lens candidates in three probability
``classes". 

%
%
\begin{figure}[tbm]
	\unitlength1cm
	\begin{picture}(14.0,8.0)   
	\put(1.5, 0.1)               
	{\epsfxsize=12.0cm           
 	\fbox{\parbox[c]{10cm}{{A Postscript version of this 
	figure can be found at \break
 	http://www.aip.de:8080/$\sim$jkw/review\_figures.html} } } 
	%
	%
	%
	}
	\end{picture}
\caption{\label{fig-lens-galaxy}
	A recent example for the identification of the lensing
	galaxy in a double quasar system \cite{Cou97}:
	The left panel shows on infrared (J-band) observation
	of the two images of double quasar HE 1104-1825
	($z_Q = 2.316$, $\Delta \theta = 3.2$ arcsec).
	The right panel obtained with some
	new deconvolution technique nicely reveals the lensing 
	galaxy (at $z_G = 1.66$) between the quasar images
	(Credits: Frederic Courbin and ESO).
}
\end{figure}

Gravitationally lensed quasar systems are studied individually 
in great detail to get a better understanding of both 
lens and source (so that, e.g.,  a measurement of the time delay
can be used to determine the Hubble constant).
As an ensemble, the lens systems 
are also analysed statistically in order to get information about 
the population of lenses (and quasars) in the universe, 
their distribution in distance (i.e. cosmic time) and mass, 
and hence about the cosmological model 
(more about that in Section \ref{sec-cosmological}).
Here we will have a close look on one particularly
well investigated system.


\subsubsection*{The first Lens: Double Quasar Q0957+561}

 The quasar Q0957+561 was originally found in a radio survey,  
 subsequently an optical counterpart was identified as well. 
 After the confirmation  of its lens nature \cite{WCW,Wal88}
 this quasar attracted quite some attention.
 Q0957+561 has been looked at in all available wavebands,
 from X-rays to radio frequencies.
 More than 100 scientific papers have appeared on Q0957+561 
 (cf. \cite{PSV}),
many more than on any other gravitational lens system. 
Here we will summarize what is known about this system from 
optical and radio observations.

In the optical light Q0957+561 appears as two point images of
roughly 17 mag  (R band) separated by 6.1 arcseconds
(see Figure \ref{Q0957+561_hst}).
The spectra of the two quasars reveal
both redshifts to be $z_Q = 1.41$. 
Between the two images, 
not quite on the connecting line, 
the lensing galaxy (with redshift $z_G = 0.36$)
appears as a fuzzy patch
close to the component B.
This galaxy is part of a cluster of galaxies at about the same
redshift. This is the reason for the relatively large separation
for a galaxy-type lens (typical galaxies with masses of 
$10^{11-12} M_{\odot}$  produce
splitting angles of only about one arcsecond, see equation 
(\ref{eq-angle-gal})). 
In this lens system the mass in the galaxy cluster helps to increase 
the deflection angles to this large separation.

%
%
\begin{figure}[tbm]
	\unitlength1cm
	\begin{picture}(14.0,10.0)   
	\put(3.0, 0.1)               
	{\epsfxsize=10.0cm           
 	\fbox{\parbox[c]{10cm}{{A Postscript version of this 
	figure can be found at \break
 	http://www.aip.de:8080/$\sim$jkw/review\_figures.html} } } 
	%
	%
	%
			}   
	\end{picture}
\caption{\label{Q0957+561_hst}
	In this 
	Hubble Space Telescope image of the 
	double quasar Q0957+561A,B.
	The two images A (bottom) and B (top) are separated by 6.1 
	arcseconds.  Image B is about 1 arcsecond away from the 
	core of the galaxy, and hence seen ``through" the halo of the 
	galaxy
        (Credits: E.E. Falco et al. -- CASTLE collaboration 
	\cite{CASTLE} -- and NASA).
}
\end{figure}

A recent image of Q0957+561 taken with the MERLIN radio telescope  
is shown in Figure \ref{Q0957+561_merlin}.
The positions of the two point-like objects in this radio
observation coincide with the optical sources. 
There is no  radio emission detected at the position of the galaxy
center, i.e. the  lensing galaxy is radio-quiet.
But this also tells us that a possible third image of the
quasar must be very faint, below the detection limit of all the
radio observations\footnote{There exists a theorem that gravitational
	lenses should produce an odd number of images 
	( e.g., \cite{McK85}).}.
In Figure \ref{Q0957+561_merlin}  a ``jet"  can be seen emerging
from image A (at the top). It is not unusual for radio quasars
to have such a ``jet" feature. This is most likely matter that is 
ejected from the central engine of the quasar with very high speed
along the polar axis of the central black hole.
The reason that this jet is seen only around one image is 
that it lies outside the caustic region in the source plane, 
which marks the part that is multiply imaged: 
only the compact core of the quasar lies inside the caustic
and is doubly imaged.

%
%
\begin{figure}[tbm]
	\unitlength1cm
	\begin{picture}(14.0,10.0)   
	\put(2.5, 0.1)               
	{\epsfxsize=10.0cm           
 	\fbox{\parbox[c]{10cm}{{A Postscript version of this 
	figure can be found at \break
 	http://www.aip.de:8080/$\sim$jkw/review\_figures.html} } } 
	%
	%
	%
			}   
	\end{picture}
\caption{\label{Q0957+561_merlin}
	Radio image of Q0957+561  from MERLIN telescope: it
	clearly shows the two point like images of the quasar
	core and the jet emanating only from the Northern part
	(Credits: N. Jackson, Jodrell Bank).
}
\end{figure}

As stated above, a virtual ``proof" of a gravitational lens
system is a measurement of the ``time delay" $\Delta t$,
the relative shift of the light curves of the two or more
images, $I_A(t)$ and $I_B(t)$, 
so that $I_B(t) = const \times I_A(t+\Delta t)$:
any intrinsic fluctuation of the quasar shows up in both
images, in general with an overall offset in apparent
magnitude and  an offset in time. 

Q0957+561 is the first lens system in which the time delay was firmly
established.  After a decade long attempt and various groups claiming 
either of two favorable values \cite{Pel96,PRH,ST95,Van89}, 
Kundi\'c et al. \cite{Kun97} 
confirmed the shorter of the two (cf. Figure \ref{fig-0957-light};
see also Oscoz et al. \cite{Osc97} and Schild \& Thomson \cite{ST97}):
\begin{equation}
\Delta t_{Q0957+561} = (417\pm3) {\rm days}
\end{equation}
%
%

%
%
\begin{figure}[tbm]
	\unitlength1cm
	\begin{picture}(14.0,11.0)   
	\put(2.5,-0.1)               
	{\epsfxsize=10.0cm           
 	\fbox{\parbox[c]{10cm}{{A Postscript version of this 
	figure can be found at \break
 	http://www.aip.de:8080/$\sim$jkw/review\_figures.html} } } 
	%
	%
	%
			}   
	\end{picture}
\caption{\label{fig-0957-light} Optical Lightcurves of images Q0957+561 
	A and B (top panel: g-band; bottom panel: r-band).  The blue 
	curve is the one of leading image A, the red one the trailing 
	image B. Note the steep drop that occured in December 1994
	in image A  and was seen in February 1996 in image B.
	The light curves are shifted in time by about 417 days 
	relative to each other
	(Credits: Tomislav Kundi\'c; see also \cite{Kun97}).
}
\end{figure}

With a model of the lens system, the time delay can be used
to determine the Hubble constant\footnote{This 
	can be seen very simply: Imagine a lens 
	situation like the one displayed in Figure \ref{fig-setup}. 
	If now all length scales are reduced by a factor of two and 
	at the same time all masses are reduced by a factor of two, 
	then for an observer the angular configuration in the sky 
	would appear exactly identical. 
	But the total length of the light path is reduced by a factor 
	of two. Now, since the time delay between the two paths is the 
	same fraction of the total lengths in either scenario, a 
	measurement of this {\bf fractional}  length allows to 
	determine the total length, and hence the Hubble constant, 
	the constant of proportionality between distance and redshift.}.
In Q0957+561 the lensing action
is done by an individual galaxy plus an associated
galaxy cluster (to which the galaxy belongs).
This provides some additional problems,
a degeneracy in the determination of the Hubble constant \cite{GFS}:
the appearance of the double quasar system including the time delay
could be identical for different partitions of the matter
between galaxy and cluster, 
but the derived value of the Hubble constant
could be quite different.
However, this degeneracy can be ``broken",
once the focussing contribution of the galaxy cluster can be determined
independently. And the latter has been attempted recently  \cite{Fis97}.
The  resulting value for the Hubble constant
\cite{Kun97} obtained by employing a 
detailed lens model 
\cite{GN} and the measured velocity dispersion of
the lensing galaxy \cite{Fal97}
is
\begin{equation}
	H_0 = (67 \pm  13) \  {\rm km} \ {\rm sec}^{-1} \ {\rm Mpc}^{-1},
\end{equation}
where the uncertainty comprises the 95\% confidence level.

	More time delays are becoming available for other  lens
	systems (e.g., \cite{Big97,Sch97a}).
	Blandford \& Kundi\'c \cite{BK97} provide a nice review in which
	they explore the potential to get a good determination of the 
	extragalactic distance scale by combining measured time 
	delays with good models;  see also 
	\cite{Sch97b}
	and \cite{WS97} for very recent summaries of the 
	current situation on time delays and determination
	of the Hubble constant from lensing.


\subsection{Quasar microlensing  }

Light bundles from ``lensed" quasars are split by 
intervening galaxies. 
With typical separations of order one arcsecond
between center of galaxy and quasar image 
this means that the quasar light bundle passes through  the
galaxy and/or the galaxy halo. 
Galaxies consist at least partly of
stars, and galaxy haloes consist possibly of compact objects as well.

Each of these stars (or other compact objects, like
black holes, brown dwarfs, or planets)
acts as a ``compact lens" or ``microlens" and produces at least one
new image of the source. 
In fact, the ``macro-image" consists of
many ``micro-images" (Figure \ref{fig-micro-image}).
But because the image splitting is proportional
to the lens mass -- see equation (\ref{eq-angle})  --
these microimages are only
of order a microarcsecond apart and can not be resolved.
Various aspects of microlensing have been addressed after the
first double quasar had been discovered
\cite{CR79,CR84,Got81,KRS,Pac86a,SW87,Wam90}.  

%
%
\begin{figure}[tbm]
	\unitlength1cm
	\begin{picture}(14.0,10.0)   
	\put(2.5,-0.1)               
	{\epsfxsize=10.0cm           
 	\fbox{\parbox[c]{10cm}{{A Postscript version of this 
	figure can be found at \break
 	http://www.aip.de:8080/$\sim$jkw/review\_figures.html} } } 
	%
	%
	%
			}   
	\end{picture}
\caption{\label{fig-micro-image}
	``Micro-Images":
	The top left panel shows an assumed ``unlensed" source
	profile of a quasar. The other three panels illustrate
	the micro-image configuration as it would be produced
	by stellar objects in the foreground. The surface mass
	density of the lenses is 20\% (top right),
	50\% (bottom left) and 80\% (bottom right) of
	the critical density (cf. equation \ref{eq-crit}),
	respectively. The angular scale of these images
	is of order microarcseconds, i.e. they cannot
	be resolved now or in the near future. Instead
	the telescopes measure the combined intensity
	of all the micro-images that form a macro-image.
}
\end{figure}

The surface mass density in front of a multiply imaged quasar
is of order the ``critical surface mass density",
see equation (\ref{eq-crit}). 
Hence microlensing should be occuring basically all the time.
This can be visualized in the following way.
If one assigns each microlens a little disk with radius equal to
the Einstein ring, then the fraction of sky which is covered
by these disks corresponds to the surface mass density
in units of the critical density; this fraction is sometimes also
called the ``optical depth".

\begin{figure}[bth]
	\unitlength1cm
	\begin{picture}(14.0,12.0)   
	\put(1.5, 0.1)               
	{\epsfxsize=12.0cm           
	%
	%
	%
	%
	%
	%
		\epsfbox{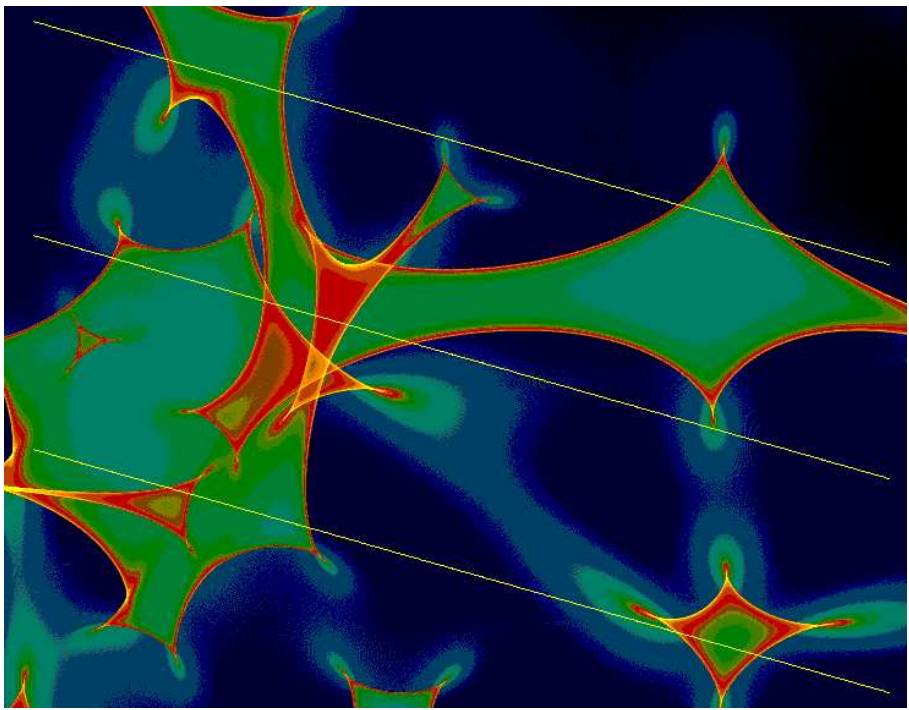}                          
			}   
	\end{picture}
\caption{\label{fig-magpat}
	Magnification pattern in the source plane, produced
	by a dense field of stars in the lensing galaxy.
	The grey tone reflects the magnification as a function of the
	quasar position: the sequence black-grey-white 
	indicates increasing magnification. Lightcurves taken
	along the white tracks are shown in Figure
	\ref{fig-lightcurve}. The microlensing parameters
	were chosen according to a model for image A of the
	quadruple quasar Q2237+0305: $\kappa = 0.36$, $\gamma = 0.44$.
}
\end{figure}

The microlenses produce a complicated two-dimensional magnification 
distribution in the source plane.
It  consists of many caustics, locations
that correspond to formally infinitely high magnification. 
An example for such a magnification pattern is shown in 
Figure  \ref{fig-magpat}.
It is determined with the parameters
of image A of the quadruple quasar Q2237+0305  (surface mass
density $\kappa = 0.36$; external shear $\gamma = 0.44$).
The grey scale indicates the magnification: dark grey is relatively 
low magnification, and white is very high magnification.

\begin{figure}[bth]
	\unitlength1cm
	\begin{picture}(14.0,10.0)   
	\put(2.5,-0.5)               
	{\epsfxsize=10.0cm           
		\epsfbox{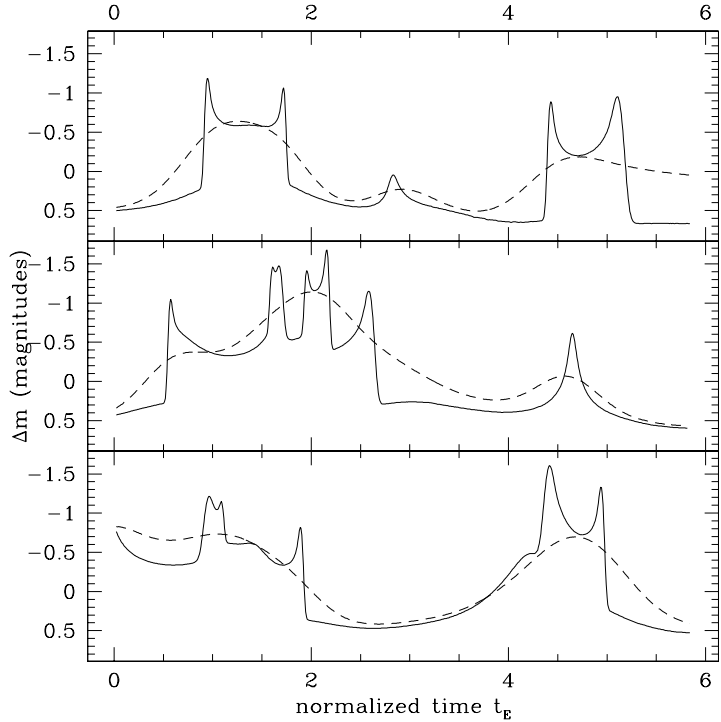}                          
			}   
	\end{picture}
\caption{\label{fig-lightcurve}
	Microlensing Lightcurve  for the white tracks in
	Figure \ref{fig-magpat}. The solid and dashed lines
	indicate relatively small and large quasar sizes. 
	The time axis is in units  of Einstein radii divided
	by unit velocity. 
}
\end{figure}

Due to the relative motion between observer, lens and source 
the quasar changes its position relative to this arrangement
of caustics, i.e.
the apparent brightness of the quasar changes with time. 
A one-dimensional cut through such a magnification pattern, convolved
with a source profile of the quasar, results in 
a microlensed lightcurve. 
Examples for microlensed lightcurves taken along the white tracks
in Figure \ref{fig-magpat}
can be seen in Figure \ref{fig-lightcurve} for two different
quasar sizes.

In particular when the quasar track crosses a caustic (the sharp lines 
in Figure \ref{fig-magpat} for which the magnification
formally is infinite, because the determinant of the Jacobian
disappears, cf. Equation (\ref{eq-magn-det})),
a pair of highly magnified microimages appears  newly or merges
and disappears (see \cite{Bla86}). 
Such a microlensing event
can easily be detected as a strong peak in the 
lightcurve of the quasar image.

%
%
%
%
%
%
%
%
%
In most simulations it is assumed that the relative
positions of the microlenses is fixed and the lightcurves
are produced only  by the bulk motion between quasar, galaxy
and observer. A visualization of a situation with changing
microlens positions 
for three different values of the surface mass density
can be found  in three video sequences
accompanying \cite{WK95}.
%
This  change of caustics  shapes due to the 
motion of individual stars produces additional
fluctuations in the lightcurve \cite{KW93,WK95}.

Microlens-induced  fluctuations in the observed brightness of quasars
contain information both about the light-emitting source
(size of continuum region or broad line region of the quasar,
brightness profile of quasar) and about
the lensing objects (masses, density, transverse velocity).
Hence from a comparison between observed and simulated quasar 
microlensing (or lack of it)
one can draw conclusions about the density and
mass scale of the microlenses.
It is not trivial, though, to extract this information quantitatively.
The reason is that in this regime of optical depth of order
one, the magnification is not due to a single isolated microlens,
but it rather  is a collective effect of many stars. 
This means individual mass determinations are not just impossible 
from the detection of a single caustic-crossing microlensing event, 
but it does not even make sense to try do so, since these events
are not produced by individual lenses\footnote{
	Similarly, one cannot determine the temperature
	of a black body by measuring the energy of a single
	photon emitted by the black body, but one needs
	to measure a large number of them and compare with
	some underlying theory.}.
Mass determinations can only be done in a statistical sense, by 
comparing good observations (frequently sampled, high photometric
accuracy) with simulations.
Interpreting microlensed lightcurves of multiple quasars allows to 
determine the size of the continuum emitting region of the quasar
and to learn even more about the central engine 
\cite{Gou97,Jar92,RB91,WPS90}. 

\begin{figure}[tbh]
	\unitlength1cm
	\begin{picture}(14.0,5.0)   
	\put(2.0,-0.1)               
	{\epsfxsize=11.0cm           
	%
	%
	%
	%
	%
	%
		\epsfbox{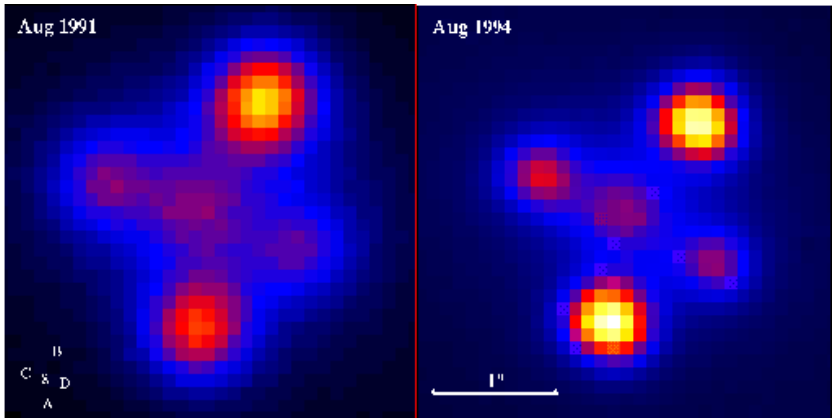}                          
			}   
	\end{picture}
\caption{\label{fig-2237-lewis}
	Two images of the quadruple quasar Q2237+0305
	separated by three years. It is obvious that
	the relative brightnesses of the images change: Image
	B is clearly the brightest one in the left panel,
	whereas images A and B are about equally bright
	in the right panel
	(Credits: Geraint Lewis).
}
\end{figure}

\begin{figure}[bth]
	\unitlength1cm
	\begin{picture}(14.0,8.0)   
	\put(2.0,-0.1)               
	{\epsfxsize=12.0cm           
	%
	%
	%
	%
	%
	%
		\epsfbox{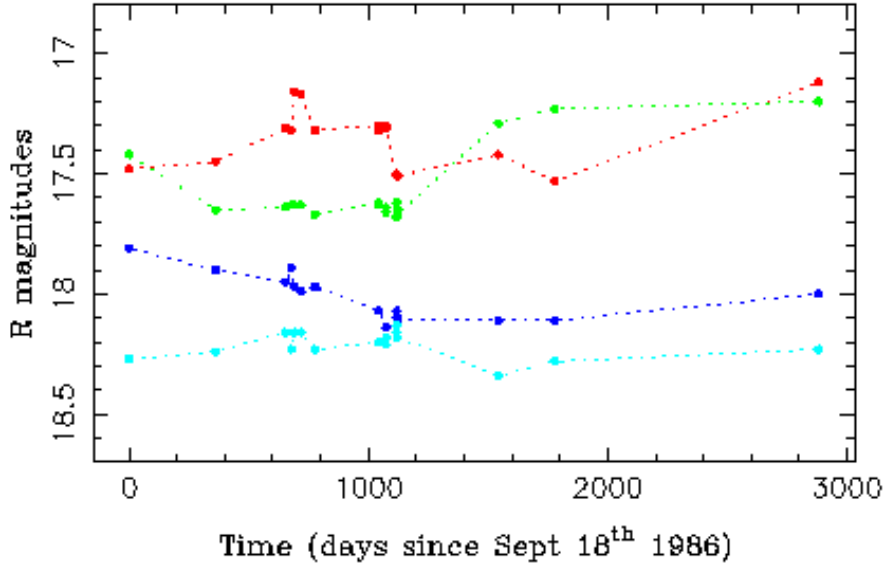}                          
			}   
	\end{picture}
\caption{\label{fig-2237-light}
	Lightcurves of the four images of Q2237+0305 over
	a period of almost ten years. The changes in relative
	brightness are very obvious
	(Credits: Geraint Lewis).
}
\end{figure}

So far the ``best" example of a microlensed quasar is
the quadruple quasar Q2237+0305 
\cite{H85,Irw89,Lew97,Ost96,WPS90,Web91,Wit93}.
In Figure \ref{fig-2237-lewis} two images of
this system are shown which were taken in 1991 and 1994, respectively. 
Whereas on the
earlier observation image B (top) is clearly the brightest,
three years later image A (bottom) is at least comparable 
in brightness. Since the time delay in this system is
only a day or shorter (because of the symmetric image arrangement),
any brightness change on larger time scales must be due to microlensing.
In Figure \ref{fig-2237-light} lightcurves are shown for the
four images of Q2237+0305 over a period of almost a decade
(from \cite{Lew95}).
The changes of the relative
brightnesses of these images induced by microlensing are obvious.

%
%
%
%
%


\subsection{Einstein rings }
\label{sec-einstein_rings}

If a point source lies exactly behind a point lens, a
ring-like image occurs.
Theorists had recognized early on \cite{Chw24,Ein36}
that such a symmetric
lensing arrangement would result in a ring-image, a 
so-called ``Einstein-ring".  Can we observe Einstein rings?
There are two necessary requirements for their occurence: 
the mass distribution of the lens needs to be axially symmetric, 
as seen from the observer,
and the source must lie exactly on top of the resulting 
degenerate point-like caustic. 
Such a geometric arrangement is highly unlikely for point-like sources. 
But astrophysical  sources in the real universe have a finite extent, 
and it is enough if a part of the source covers the point caustic 
(or the complete astroid caustic in
a case of  a not quite axial-symmetric mass distribution)
in order to produce such an annular image.

In 1988 the first example of an ``Einstein ring" was discovered
\cite{Hew88}. With high resolution radio observations,
the extended radio source  MG1131+0456 turned out 
to be a ring with a diameter of about 1.75 arcsec.  
The source was identified as a radio lobe at a redshift 
of $z_S = 1.13$, whereas the lens is a galaxy at $z_L = 0.85$.
Recently a  remarkable observation of the Einstein ring 1938+666
was presented \cite{Kin97}. The infrared HST image shows an almost
perfectly circular ring with two bright parts plus the bright central
galaxy. The contours agree very well with the MERLIN radio map
(see Figure \ref{fig-1938}).

%
%
\begin{figure}[hbt]
	\unitlength1cm
	\begin{picture}(14.0,10.0)   
	\put(1.5, 0.1)               
	{\epsfxsize=12.0cm           
 	\fbox{\parbox[c]{10cm}{{A Postscript version of this 
	figure can be found at \break
 	http://www.aip.de:8080/$\sim$jkw/review\_figures.html} } } 
	%
	%
	%
			}   
	\end{picture}
\caption{\label{fig-1938}
	Einstein ring 1938+666 (from \cite{Kin97}): The left
	panel shows the radio map as contour superimposed on the
	grey scale HST/NICMOS image; the right panel is a 
	b/w 
	depiction of the infrard HST/NICMOS image. The diameter
	of the ring is about 0.95 arcseconds
	(Credits: Neal Jackson and NASA).
}
\end{figure}

By now about a half dozen cases have been found that qualify
as Einstein rings.
Their diameters vary between 0.33 and about 2 arcseconds.
All of them are found in the radio regime, some have optical
or infrared counterparts as well.
Some of the Einstein rings are not really complete rings, but they 
are ``broken" rings with  one or two interruptions along the circle. 
The sources of most Einstein rings have both an extended and a compact
component. The latter is always seen as a double image, separated
by roughly the diameter of the Einstein ring.  In some cases
monitoring of the radio flux showed that the compact
source is variable. This gives the opportunity to
measure the time delay and the Hubble constant $H_0$
in these systems.  

The Einstein ring systems provide some advantages over the
multiply-imaged
quasar systems for the goal to determine the
lens structure and/or the Hubble constant. 
First of all the extended image structure 
provides many constraints on the lens. A lens  model can be
much better determined than in cases of just two or three
or four point-like quasar images.
Einstein rings thus help us to understand the mass distribution
of galaxies at moderate redshifts.
For the Einstein ring MG 1654+561 it was found \cite{Koc95} 
that the radially averaged surface mass density of the lens 
was fitted well with a  distribution like
$\Sigma(r) \propto r^{\alpha}$,  
where $\alpha$ lies between $-1.1 \le \alpha \le -0.9$ 
(an isothermal sphere would have exactly $\alpha = -1$!); there
was also evidence found for dark matter in this lensing galaxy.

Second, since the diameters of the observed rings (or the separations 
of the accompanying double images) are of order
one or two arcseconds,  the expected time delay must be  much
shorter than the one in the double quasar Q0957+561 (in fact, it can
be arbitrarily short, if the source happens
to be very close to the point caustic). This means one does not
have to wait so long to establish a time delay (but the source
has to be variable intrinsically on even shorter time scales ...).

The third advantage is that since the emitting region of the
radio flux is presumably much larger than that of the optical
continuum flux, the radio lightcurves of the different
images are not affected by microlensing. Hence the radio lightcurves
between the images should agree with each other very well.

Another interesting application is the (non-)detection of
a central image in the Einstein rings. For singular lenses, 
there should be no central image (the reason is the
discontinuity of the deflection angle).
However, many galaxy models
predict a finite core in the mass distribution of a galaxy. 
The non-detection of the central images puts strong
constraints on the size of the core radii.


\subsection{Giant luminous arcs and arclets }
\label{sec-arcs}

Zwicky had pointed out the potential use in the 1930s, 
but nobody had really followed up the idea, not even 
after the discovery of the lensed quasars:
Galaxies can be gravitationally lensed as well. 
Since galaxies are extended objects, the apparent consequences 
for them would be far more dramatic than for quasars: galaxies
should be heavily deformed once they are strongly lensed.

It came as quite a surprise when in 1986 
Lynds \& Petrosian \cite{LP86} and Soucail et al. \cite{Sou87}
independently discovered this new gravitational lensing phenomenon: 
magnified, distorted and strongly elongated images
of background galaxies which happen to lie behind 
foreground clusters of galaxies.

Rich clusters of galaxies at redshifts beyond $z \approx 0.2$ 
with masses of  order $10^{14} M_{\odot}$ are very effective
lenses if they are centrally concentrated.
Their Einstein radii are of the order of 20 arcseconds. 
Since most clusters are not really spherical mass distributions
and since the alignment between lens and source is usually not perfect,
no complete Einstein rings have been found around clusters.
But there are many examples known with spectacularly long arcs which
are curved around the cluster center, with lengths up to about 20 arcseconds.

The giant arcs can be exploited in two ways, as is typical
for many lens phenomena. Firstly they provide us with strongly magnified
galaxies at (very) high redshifts. These galaxies would be too faint
to be detected or analysed in their unlensed state. 
Hence with the lensing boost we can study these 
galaxies in their
early evolutionary stages, possibly as infant or proto-galaxies,
relatively shortly after the big bang. 
The other practical application of the arcs is to take them as tools to 
study the potential and  mass distribution of the lensing galaxy 
cluster. 
In the simplest model of a spherically symmetric mass distribution
for the cluster, giant arcs form very close to the critical curve,
which marks the Einstein ring. So with the redshifts of
the cluster and the arc it is easy to determine a rough
estimate of the lensing mass  by just determining the radius of
curvature and interpreting it as the Einstein radius of the lens system.

More detailed modelling of the lensing clusters which allows for 
the asymmetry of the mass distribution according to the visible
galaxies plus an unknown dark matter component provides more
accurate determinations for the total cluster mass and its 
exact distribution. More than once this detailed
modelling predicted additional (counter-) images of giant arcs, 
which later were found and confirmed spectroscopically
\cite{Kne93,Ebb97}.

Gravitational lensing is the third method for the 
determination of masses of galaxy clusters,
complementary to the mass determinations
by X-ray analysis and the old art of using the virial theorem and
the velocity distribution of the galaxies (the latter two methods
use assumptions of hydrostatic or virial equilibrium, respectively).
Although there are still some discrepancies between the 
three methods, it appears that in
relaxed galaxy clusters the agreement between these different
mass determinations is very good  \cite{All97}.

Some general results from the analysis of giant arcs in galaxy
clusters are: 
Clusters of galaxies are dominated by dark matter. 
The typical ``mass-to-light ratios"  
for clusters obtained from strong (and weak, see below) lensing 
analyses are \ \ $M/L \ge 100 M_\odot/L_\odot$.
The distribution of the dark matter follows roughly the 
distribution of the light in the galaxies, 
in particular in the central part of the cluster.
The fact that we see such arcs shows that the central surface
mass density in clusters must be high. 
The radii of curvature of many giant arcs is comparable to
their distance to the cluster centers; this shows that 
core radii of clusters --
the radii at which the mass profile of the cluster 
flattens towards the center -- must be of order
this distance or smaller.
For stronger constraints detailed modelling of the mass
distribution is required.

\begin{figure}[hbt]
	\unitlength1cm
	\begin{picture}(14.0,8.0)   
	\put(1.5, 0.1)               
	{\epsfxsize=12.0cm           
	%
	%
	%
	%
	%
	%
		\epsfbox{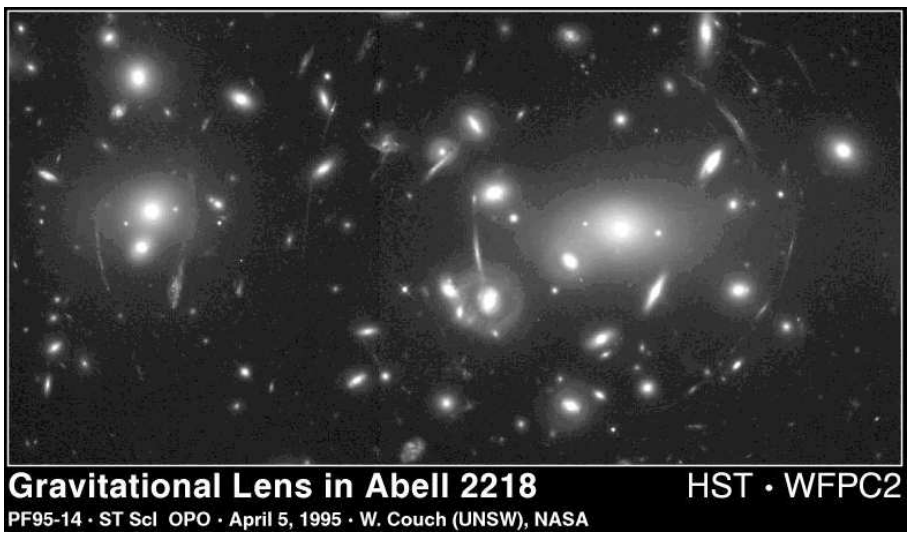}                          
			}   
	\end{picture}
\caption{\label{fig-2218} Galaxy Cluster Abell 2218 with Giant 
	Luminous Arcs and 
	many arclets,  imaged with the Hubble Space Telescope.
	The original picture (and more information)
	can be found 
	in \cite{Kne96}
	(Credits: W.Couch, R. Ellis and NASA).
}
\end{figure}

\begin{figure}[bth]
	\unitlength1cm
	\begin{picture}(14.0,13.0)   
	\put(1.5, 0.1)               
	{\epsfxsize=12.0cm           
	%
	%
	%
	%
	%
	%
		\epsfbox{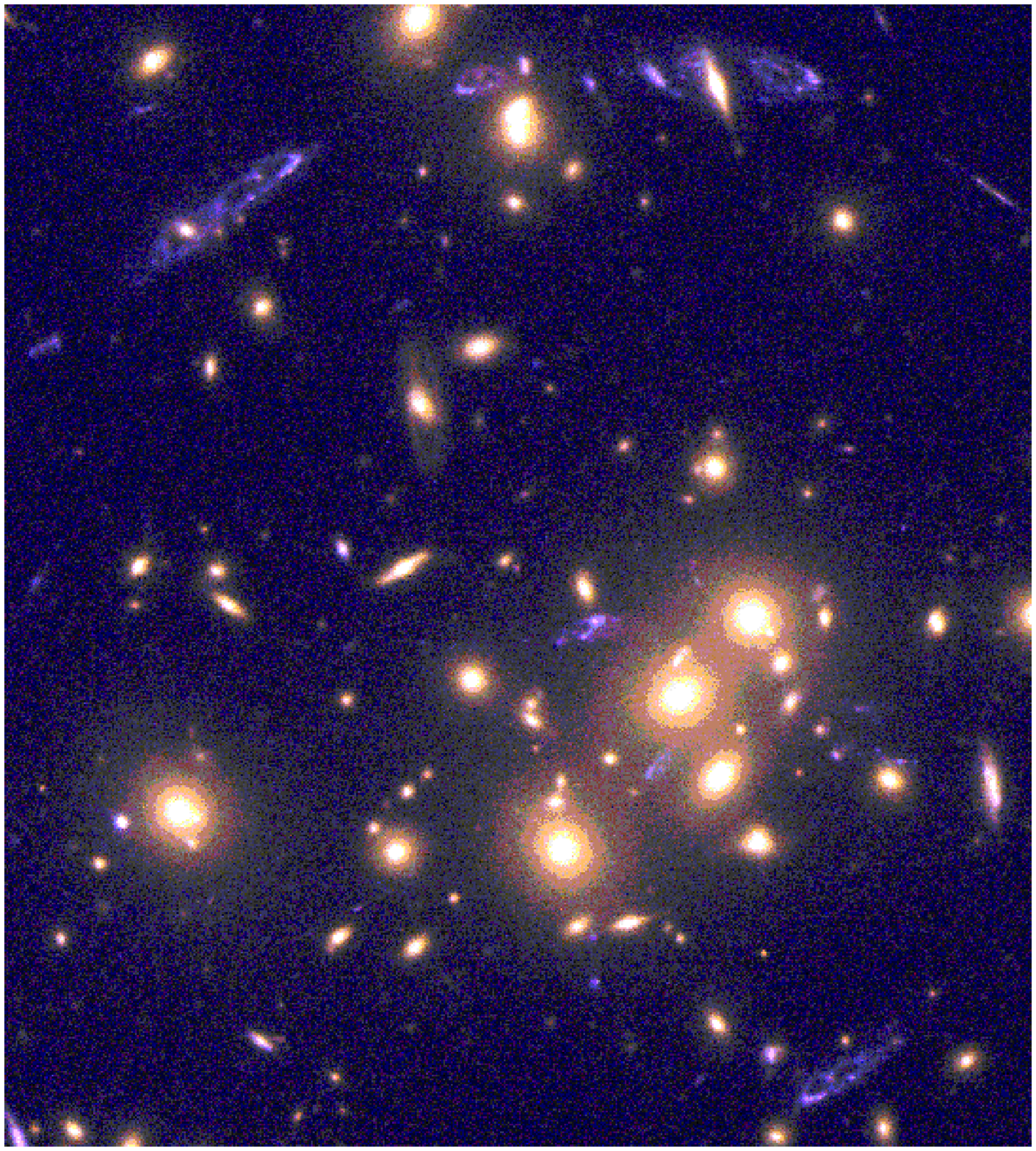}                          
			}   
	\end{picture}
\caption{\label{fig-0024}
	Galaxy Cluster CL0024+1654 with multiple images of  a 
	blue background galaxy.
	A scientific analysis which includes a reconstruction
	of the source galaxy can be found in \cite{CTT96}
	(Credits: W.N. Colley, E. Turner,
	J.A. Tyson and NASA).
	}
\end{figure}

In Figures \ref{fig-2218} and \ref{fig-0024}
two of the most spectacular cluster lenses producing arcs
can be seen:
Clusters Abell 2218
and  
CL0024+1654.
Close inspection of the HST image of Abell 2218 reveals that the
giant arcs are resolved (Figure \ref{fig-2218}),  
structure can be seen in the individual
components \cite{Kne96} 
and used for detailed mass models of the lensing cluster.
In addition to the giant arcs, 
more than 100 smaller ``arclets" can be identified in Abell 2218.
They are farther away from the lens center and hence are not 
magnified and curved as much as the few giant arcs. 
These arclets are all slightly distorted images of background galaxies.
With the cluster mass model it is possible to predict
the redshift distribution of these galaxies. This
has been successfully done in this system with the identification
of an arc as a star-forming region, opening
up a whole new branch for the application of cluster lenses
\cite{Ebb96}.

In another impressive exposure with the Hubble Space Telescope, 
the galaxy cluster CL0024+1654 (redshift $z = 0.39$) 
was deeply imaged in two filters \cite{CTT96}. 
The combined picture (Figure \ref{fig-0024}) shows very 
nicely the reddish images of cluster galaxies, the
brightest of them concentrated around the center, and the
bluish arcs.
There are four blue images which all have a shape 
reminiscent of the Greek letter $\Theta$.
All the images are resolved and show similar structure
(e.g., the bright fishhook-like feature at one end
of the arcs), but two of them are mirror inverted, i.e. have
different parity! 
They lie roughly on a circle around the center of
the cluster and are tangentially elongated. 
There is also another  faint blue image  relatively close to the
cluster center, which is extended radially. 
Modelling reveals that this
is a five-image configuration produced by
the massive galaxy cluster. 
All the five arcs are  images of
the same galaxy, which is far behind the cluster at a  much higher
redshift and most likely undergoes a burst of star formation.
This is a spectacular example  of the use of a galaxy 
cluster as a ``Zwicky" telescope.

%
%

%
%

In CL0024+1654 the lensing
effect produces a magnification of roughly a factor of ten.
Combined with the angular resolution of the HST of 0.1 arcsec,
this 
can be used to yield a resolution that effectively
corresponds to 0.01 arcsec (in the tangential direction), 
unprecedented in direct optical imaging.
Colley et al.  \cite{CTT96} map
the five images ``backward" to the source plane
with their model for the cluster lens and hence reconstruct
the un-lensed source. They get basically 
identical source morphology for all arcs, which confirms
that the arcs are all images of one source.

Recently, yet another superlative about cluster lenses was found:
A new giant luminous arc was discovered  in the field
of the galaxy cluster CL1358+62 with the HST \cite{Fra97}.
This arc-image  turned out to be a galaxy at a redshift of $z = 4.92$. 
Up to a few months ago this was
	the  most distant object in the universe with a
	spectroscopically measured redshift!
In contrast to most other arcs, this one is very red. 
The reason is that due to this very high redshift, the
Lyman-$\alpha$ emission of the galaxy, which is emitted
in the ultra-violet part of the electromagnetic spectrum at a wavelength
of  1216 \AA \ is shifted
by a factor of $z+1 \approx 6$ to the red part at a wavelength
of about $7200$\AA!

A review of cluster lensing and giant arcs/arclets can
be found in Fort \&  Mellier \cite{FM94}.
The review by Wu \cite{Wu96} provides, e.g., an updated
table of giant arcs.


\subsection{Weak/statistical lensing }

In contrast to the phenomena that were mentioned
so far,  ``weak lensing" 
deals with effects of light deflection that cannot
be measured individually, but rather in a statistical way only.
As was discussed above
``strong lensing" -- 
usually defined as  the regime that involves multiple images, high
magnifications, and caustics in the source plane --
is a rare phenomenon.
Weak lensing on the other hand is much more common.
In principle, weak lensing acts along each line of sight in the 
universe, since each photon's path is affected
by matter inhomogeneities along or near its path.
It is just a matter of how accurate we can measure (cf. \cite{Pre95}).

Any non-uniform matter distribution between our
      observing point and     distant light sources
affects the measurable properties of the sources 
in two different ways:
the angular size of extended objects is changed 
and the apparent brightness of a source is affected, as was
first formulated in 1967 by Gunn \cite{Gun67a,Gun67b}.

A weak lensing effect can be a small 
deformation of the shape of a cosmic object,
or a small modification of its brightness,
or a small change of its position.  
In general the latter cannot be observed, 
since we have no way 
of knowing the unaffected position\footnote{A well-known exception is 
the light deflection at the solar limb,  where the difference between
the lensed and the unlensed positions of stars was
used to confirm General Relativity, see Chapter \ref{sec-hist}.}.

The first two effects -- slight shape deformation or small change
in brightness -- in general cannot be determined
for an individual image. Only when averaging over a 
whole ensemble of images
it is  possible to measure the shape distortion,
since the weak lensing (due to mass distributions
of large angular size) acts as the coherent deformation 
of the shapes of extended background sources.

The effect on the apparent brightness of sources
shows that weak lensing can be both a blessing and a curse for 
astronomers: 
the statistical incoherent lens-induced change of the
apparent brightness of (widely separated)
``standard candles" -- like type Ia supernovae --
affects the accuracy of the determination
of cosmological parameters \cite{Fri97,Kan95,Wam97}.

The idea to use the weak distortion and tangential alignment of
faint background galaxies to map the mass distribution
of foreground galaxies  and clusters
has been floating around for a long time.
The first attempts go back to the years 1978/79, 
when Tyson and his group tried to measure the
positions and orientations of the then newly discovered faint blue
galaxies, which were suspected to be at large distances.
Due to the not quite adequate techniques at the time 
(photographic plates), these efforts ended unsuccessfully
\cite{VJMT84,VTJ83}.  
Even with the advent of the new technology of CCD cameras, it
was not immediately possible to detect weak lensing, since
the pixel size originally was relatively large
(of order an arcsecond). Only with smaller CCD pixels,
improved seeing conditions at the telecope sites and
improved image quality of the telescope optics  the
weak lensing effect could ultimately be measured. 

Weak lensing is one of the two sub-disciplines within the
field of gravitational lensing with the highest rate of growth
in the last couple of years (along with galactic microlensing).
There are a number of reasons for that:
\renewcommand{\labelenumi}{\alph{enumi})}
\begin{enumerate}
\item  the availability of astronomical sites with very good 
	seeing conditions,
\item the availability of large high resolution cameras with fields
	of view of half a degree at the moment (aiming for more),
\item  the availability of methods to analyse these coherent small 
	distortions 	
\item the awareness of both observers and time allocation 
	committees about the potential of these weak lensing 
	analyses for extragalactic research and cosmology.
\end{enumerate}
\renewcommand{\labelenumi}{\arabic{enumi})}
Now we will briefly summarize the technique of how to use the weak
lensing distortion in order to get the mass distribution of
the underlying matter.


\subsubsection*{Cluster mass reconstruction}

The first real detection of  a coherent weak lensing signal
of distorted background galaxies 
was measured in 1990 
around
the galaxy clusters Abell 1689 and CL1409+52  \cite{TVW90}.
It was shown that the orientation of background galaxies -- 
the angle of the semi-major axes of the elliptical isophotes 
relative to the center of the cluster -- was  
more likely to be tangentially oriented relative to the
cluster than radially. For an unaffected population
of background galaxies one would expect no preferential direction.
This analysis is based on the assumption that the major axes 
of the background galaxies are intrinsically randomly oriented.

With the elegant and powerful method developed by Kaiser and 
Squires \cite{KS93} 
the weak lensing signal can be used to quantitatively reconstruct the 
surface mass distribution of the cluster. 
This method relies on the fact that 
the convergence $\kappa (\theta)$ and the two components of the 
shear $\gamma_1(\theta)$, $\gamma_2(\theta)$ 
are linear combinations of the 
second derivative of the effective lensing potential
$\Psi(\theta)$
(cf. equations \ref{eq-kappa} - \ref{eq-gamma2}).
After Fourier transforming the expressions for the convergence
and the shear one obtains linear relations between the transformed
components $\tilde \kappa$, $\tilde \gamma_1$, $\tilde \gamma_2$.
Solving for $\tilde \kappa$ and inverse Fourier transforming
gives an estimate for the convergence $\kappa$
(details can be found in  \cite{KS93}, \cite{Kai95b},
\cite{NB96}, or \cite{SK96}).

The original Kaiser-Squires method was 
improved/modified/extended/generalized 
by various authors subsequently. In particular the
constraining  fact that observational data are available
only in a relatively small, finite area was implemented.
Maximum likelihood techniques, non-linear reconstructions
as well as methods using the amplification effect rather 
than the distortion effect complement each other.
Various variants of the mass reconstruction technique have
been successfully applied
to more than a dozen rich clusters by now. 
Descriptions of various techniques
and applications for the cluster mass reconstruction can
be found in, e.g.,
\cite{Abd97, Bar95, Bar96a, Bar96, Bon95, 
Bro95, Ham97, Kai95a, Nat96, SS95, Sei96, SS96, 
Sma95, Wil96}.
In Figure \ref{fig-hoekstra} a recent example for the reconstructed
mass distribution of galaxy cluster CL1358+62 is shown \cite{Hoe97}.

%
%
\begin{figure}[bth]
	\unitlength1cm
	\begin{picture}(14.0,12.3)   
	\put(1.5, -0.1)               
	{\epsfxsize=12.0cm           
 	\fbox{\parbox[c]{10cm}{{A Postscript version of this 
	figure can be found at \break
 	http://www.aip.de:8080/$\sim$jkw/review\_figures.html} } } 
	%
	%
	%
			}   
	\end{picture}
\caption{\label{fig-hoekstra}
The reconstructed mass distribution of cluster
CL1358+62 from a weak lensing analysis is shown as
contour lines superposed on the image taken with the
Hubble Space Telescope \cite{Hoe97}. The map is smoothed
with a Gaussian of size 24 arsec (see shaded circle).
The center of the mass distribution agrees
with the central elliptical galaxy. 
The numbers indicate the reconstructed surface mass density in
units of the critical one
(Credits: Henk Hoekstra).
}
\end{figure}

We could present here only one weak lensing issue in some detail:
the reconstruction of the mass distribution of galaxy clusters
from weakly distorted background images. 
Many more interesting weak lensing applications are under 
theoretical and observational investigation, though.
To name just a few:
\begin{itemize}
	\item constraints on the distribution of the faint galaxies from
		weak lensing (e.g., \cite{FM97,LK97});
	\item galaxy-galaxy lensing (e.g., \cite{Bra96});
	\item lensing by galaxy halos in clusters (e.g., \cite{Nat97});
	\item weak lensing effect by large scale structure and/or
		detection of dark matter concentrations; this considers
		both shear effects as well as magnification
		effects (e.g. 
		\cite{barkana,Bar95a,Sch96,Wil97});
	\item determination of the power spectrum of the matter 
		distribution (e.g., \cite{Ber97}; or 
	\item the weak lensing effects on the cosmic microwave 
		background (e.g. \cite{Mar97}, \cite{Met97b}, 
		\cite{Sel96}).
\end{itemize}
An upcoming comprehensive
review on weak lensing by Schneider \& Bartelmann \cite{SB98}
treats both theory and applications of weak lensing in  great depths.



\subsection{Cosmological aspects of (strong) lensing }
\label{sec-cosmological}

Gravitational lenses can be used in two different ways
to study the cosmological parameters of the universe.
The first is to explore a particular lens system in great detail, 
determine all possible observational parameters 
(image positions/brightnesses/shapes; matter/light distribution of lens; 
time variability etc.) and model
both lens and source in as much detail as possible.
This way one can in principle determine the amount of dark matter 
in the lens and -- maybe even more importantly --
the value of the Hubble constant. A reliable determination of the
Hubble constant establishes the extragalactic distance scale, something
astronomers have been trying to do for more than 70 years.

The second approach is of statistical nature: find out {\it how many}
(what fraction of) quasars are multiply imaged by gravitationally 
lensing, determine their separation and
redshift distributions \cite{TOG} and deduce the value of (or limits to)
$\Omega_{\rm compact}$ -- matter in clumps of, say, 
$10^6 \le M/M_{\odot} \le 10^{14}$ --  and to 
$\Omega_{\Lambda}$ -- the value of the cosmological constant.

The  first approach has already been treated in Section \ref{sec-mult}.
Here we will concentrate on the statistical approach. 
In order to determine which fraction of a certain group of
objects is affected by strong lensing (i.e. multiply imaged), 
one first needs a well-defined underlying sample.
What needs to be done is the following:

\begin{enumerate}
   \item Do a systematic study of a sample of high-redshift objects:
	quasar surveys.
   \item Identify the gravitational lens systems among them.
   \item Determine the relative frequency of lensed objects,
		the distribution of splitting angles $\Delta \theta$ 
		as a function of  lens and source 
		redshifts $z_L$/$z_S$.
    \item Determine matter content of universe $\Omega_{\rm compact}$, 
	 typical mass scale $M_{lens}$,
   	 cosmological constant $\Omega_\Lambda$, 
	by comparison with theoretical models/simulations.
\end{enumerate}

Since quasars are rare objects and 
lensing is a relatively rare phenomenon, steps 1 and 2 are
quite difficult and time-consuming. 
Nevertheless, a number of systematic quasar surveys with the 
goal to find (many) lens systems with well defined 
selection criteria
have been done in the past and others are underway right now
(e.g., \cite{Bro97, Mao93, Mao97, Web88, Yee93}).

The largest survey so far, the CLASS survey, 
has looked at about 7000 radio sources at the moment (the goal
is 10000).
In total CLASS found 12 new lens systems so far. 
Interestingly, all the lenses have small separations 
($\Delta \theta < 3$arcsec), and all lensing galaxies are 
detected \cite{Bro97,Jac97a}. 
That leaves little space for a population of dark objects
with masses of galaxies or beyond.
A detailed discussion of lens surveys and a comparison between 
optical and radio surveys can be found in \cite{Koc96a}.

The idea for the determination of the cosmological constant
$\Omega_{\Lambda} = \Lambda/(3 H_0^2)$
from lens statistics is based on the fact that the relative
lens probability  for multiple imaging increases rapidly with 
increasing $\Omega_{\Lambda}$ (cf. Figure 9 of \cite{CPT92}).
This was first pointed out 1990 \cite{FFK90, Tur90}.
The reason 
is the fact that the angular diameter distances $D_S$, $D_L$, 
$D_{LS}$ depend strongly on the cosmological model.
And the properties that determine the probability for 
multiple lensing (i.e. the ``fractional volume" that is affected
by a certain lens) depend on these distances \cite{CPT92}.
This can be seen, e.g.,  when one looks at the 
critical surface mass density required for multiple imaging
(cf. equation \ref{eq-crit}) which depends on the angular diameter
distances.

The consequences of lensing studies
on the cosmological constant can be summarized as follows.
The analyses of the frequency of lensing are based on lens systems 
found in different optical and radio surveys. 
The main problem is still the small number of lenses. 
Depending on the exact selection criteria, only a few lens
systems can be included in the analyses.
Nevertheless, one can use the existing samples to 
put limits on the cosmological constant. Two different
studies found 95\%-confidence limits of 
	$\Omega_{\Lambda} < 0.66$ \cite{Koc96b} 
	and 
	$\Omega_{\Lambda} < 0.7$ \cite{MR93, Rix96}.
This is based on the assumption of a flat universe
	($\Omega_{matter} + \Omega_{\Lambda} = 1$).
Investigations on the matter content of the 
universe from (both ``macro-" and ``micro-") lensing
generally conclude that the
fractional matter in compact form cannot exceed a few
percent of the critical density (e.g. \cite{Can82,Dal94,Nem91,Sch93}).


\subsection{Galactic microlensing}
\label{sec-gal-micro}

It has been known for more than two decades that halos of galaxies
must contain some unknown kind of dark matter.  
Many different particles/objects had been suggested as
constituents of this halo dark matter.
The candidates can be divided into the two broad categories
``elementary particles" and ``astronomical bodies".
A conservative candidate for this dark matter are brown dwarfs,
objects with masses less than 0.08 $M_{\odot}$ so that the 
central temperature
is not high enough to start helium fusion. These objects are certain
to exist, we just do not know how many there are.

In 1986 Paczy\'nski \cite{Pac86b} suggested a method to test observationally
whether the Milky Way halo is made of such brown dwarfs 
(or other astronomical objects in roughly this mass range).
Subsequently this type of dark matter candidate was
labelled ``Macho" for MAssive Compact Halo Object \cite{Gri91}.
If one could continuously observe the brightness of stars of our
neighbouring galaxy Large Magellanic Cloud  (LMC)
one should
see typical fluctuations in some of these stars due
to the fact that every now and
then one of these compact halo objects passes in front of the
star and magnifies its brightness. 
The only problem  with this experiment
is  the low probability for such an event:
only about one out of three million LMC stars would be significantly
magnified at any given time.

The underlying scenario is very simple: 
Due to the relative motion of observer, lensing Macho and source star
the projected impact parameter between lens and source changes
with time and produces a time dependent magnification. If
the impact parameter is smaller than an Einstein radius
then the magnification 
is $\mu_{min} >  1.34$ (cf. equation \ref{eq-magn-tot}).

For an extended source such a sequence is illustrated in 
Figure  \ref{fig-macho-1}.
The separation of the two images
is of order two Einstein radii when they are of comparable
magnification, which corresponds to only about a
milliarcsecond. Hence the two images cannot be resolved
individually, 
we can only observe the brightness of the combined image pair. 
This
is illustrated in Figures \ref{fig-macho-2a} and \ref{fig-macho-2b}
which show the relative tracks and the respective light curves
for five values of the minimum impact parameter $u_{min}$.

%
%
\begin{figure}[hbt]
	\unitlength1cm
	\begin{picture}(14.0,5.0)   
	\put(1.5,-0.1)               
	{\epsfxsize=12.0cm           
 	\fbox{\parbox[c]{10cm}{{A Postscript version of this 
	figure can be found at \break
 	http://www.aip.de:8080/$\sim$jkw/review\_figures.html} } } 
	%
	%
	%
			}   
	\end{picture}
\caption{\label{fig-macho-1}
	Five snapshots of a gravitational lens situation: 
	From left to right the alignment between lens and source 
	gets better and better, until it is perfect in the rightmost 
	panel. This results in the image of an ``Einstein ring".
}
\end{figure}

%
%
\begin{figure}[hbt]
	\unitlength1cm
	\begin{picture}(14.0,9.0)   
	\put(0.5,-2.2)               
	{\epsfxsize=12.0cm           
		\epsfbox{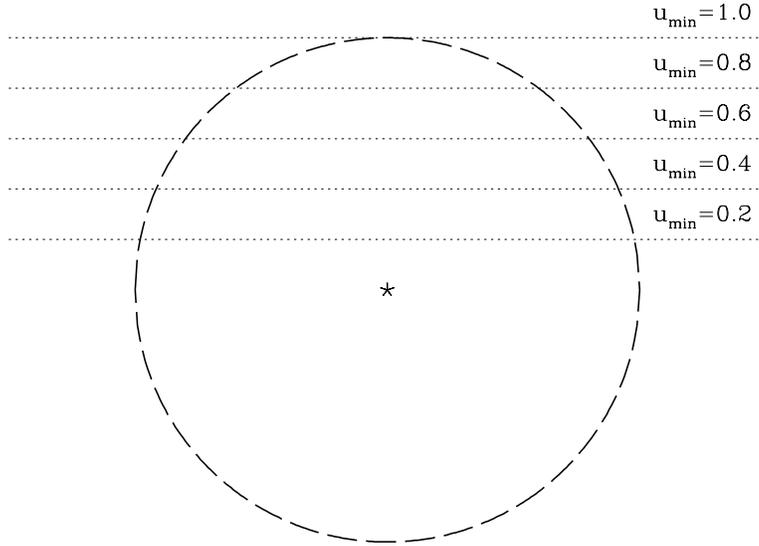}                          
			}   
	\end{picture}
\caption{\label{fig-macho-2a}
	Five relative tracks between background star and foreground lens
	(indicated as the central star) parametrized by the impact 
	parameter $u_{min}$. The dashed line indicates the Einstein ring
	for the lens 
	(after \cite{Pac86b}).
}
\end{figure}

%
%
\begin{figure}[hbt]
	\unitlength1cm
	\begin{picture}(14.0,11.0)   
	\put(1.5,-0.5)               
	{\epsfxsize=12.0cm           
		\epsfbox{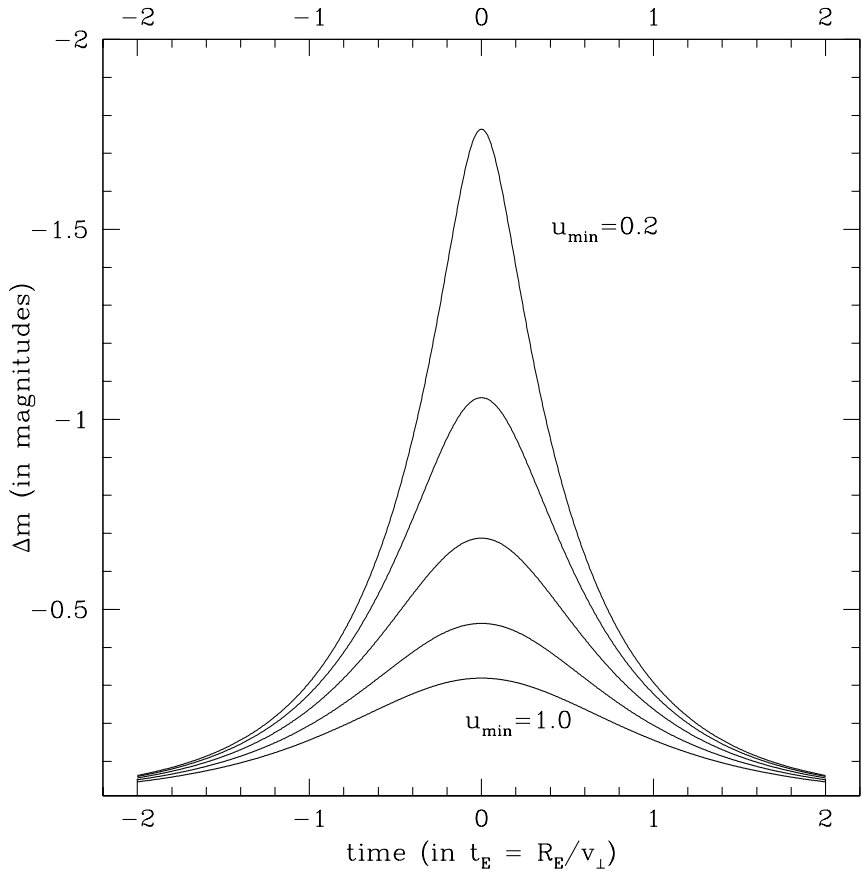}                          
			}   
	\end{picture}
\caption{\label{fig-macho-2b}
	Five microlensing lightcurves for the
	tracks indicated in Fig. \ref{fig-macho-2a}, parametrized by the
	impact parameter $u_{min}$. The verical axes is the 
	magnification in astronomical magnitudes relative to the 
	unlensed case, the horizontal axis displays the time 
	in ``normalized" units (after \cite{Pac86b}).
}
\end{figure}

Quantitatively,
the total magnification $\mu = \mu_1 + \mu_2$ of the two
images (cf. equation \ref{eq-magn-tot}) entirely
depends on the impact parameter $u(t) = r(t)/R_E$ between the lensed
star and the lensing object, measured in the lens plane
(here $R_E$ is the Einstein radius  of the lens, i.e. the
radius at which a circular image appears for perfect alignment
between source, lens and observer, cf. Figure  \ref{fig-macho-1}, 
rightmost panel):
\begin{equation}
\label{eq-light}
\mu(u) = { {u^2 + 2 } \over u \sqrt{ (u^2 + 4) } }.
\end{equation}

The time scale of such a ``microlensing event" is defined as
the time it takes the source to cross the Einstein radius:
\begin{equation}
\label{eq-macho-time}
t_0    = {R_E \over v_{\perp} } \ \approx \ \ 0.214 \ \  {\rm yr} \ \ \ 
\sqrt{M   \over M_\odot} \ 
\sqrt{D_L   \over 10 {\rm kpc}} \ 
\sqrt{1 - {D_L   \over D_S}            }  \ 
\left({v_\perp   \over 200 {\rm km/sec}}\right)^{-1}.
\end{equation}
Here $v_{\perp}$ is the (relative) transverse velocity of the 
lens. We parameterized the time scale
by  ``typical" numbers for the distances of lensed and lensing
star and the relative transverse velocity. 
Note also that here we used the simple relation
$D_{LS} = D_S - D_L$ (which is not valid for cosmological
distances).

Note that from equation (\ref{eq-macho-time}) 
it is obvious that it is {\bf not} possible
to determine the mass of the lens from one individual
microlensing event.
The duration of an event is determined by three
unknown parameters: the mass of the lens, the transverse
velocity and the distances of lens and source. It is impossible
to disentangle these for individual events. 
Only with a model for the 
spatial and velocity distribution of the lensing objects and comparison
with ``simulated microlensing events" it is possible to obtain
information about the masses of the lensing objects and their
density.

What seemed to be an impossible task at the time -- namely
determine the brightness of millions of stars on an almost
nightly basis -- 
was turned into
three big observational campaigns within few years (MACHO, EROS, OGLE
experiments). These groups looked at millions
of stars in the LMC and towards the bulge of the Milky
Way, and their first results appeared essentially
simultaneously in the fall of 1993 \cite{macho93, eros93, ogle93}.
In the meantime  more groups have joined this effort, some of
them with special emphases: 
e.g. on covering ongoing microlensing events (PLANET, DUO),
or on extending the microlensing search to unresolved
stars (``pixel lensing") in the Andromeda galaxy 
\cite{Cro96,Gou96} (AGAPE)
or to cover the Magellanic Clouds completely around the year (MOA).
Here is a list of groups currently active in the search
for microlensing signatures
of compact objects in the halo of the Milky Way or elsewhere:

\begin{itemize}
\item MACHO (MAssive Compact Halo Object): 
	\cite{macho97a,macho97b}

\item EROS   (Experience de Recherche d'Objets Sombres): 
	\cite{eros96,eros97}

\item OGLE   (Optical Gravitational Lens Experiment): 
	\cite{ogle94a,ogle97}

\item AGAPE  (Andromeda Galaxy and Amplified Pixels Experiment):
	\cite{agape97}

\item MOA (MACHO Observations in Astrophysics):
	\cite{moa96}

\item PLANET (Probing Lensing Anomalies NETwork):
	\cite{planet96}

\item DUO (Disk Unseen Objects) \cite{Ala97}

\item GMAN (Global Microlensing Alert Network)
	\cite{gman97}

\end{itemize}

The observations towards the Large 
Magellanic Cloud show that there are fewer microlensing events
than one would expect if the halo of the Milky Way was made
entirely of these compact objects.
The latest published results from the microlensing experiments that
monitor stars in the LMC 
indicate that  the optical depths toward the LMC is about
$\tau \approx 3 \times 10^{-7}$.
The observations
are consistent with 50\% of the Milky Way  halo 
made of compact objects  with most likely 
masses of $0.5^{+0.3}_{-0.2}  M_{\odot}$ \cite{macho97a}.
But the number of observed events is still small (in this
analysis eight events were used) and hence the
uncertainties are large; in fact, it cannot even be excluded
that none of the observed events is due to an unknown halo
population \cite{Gat97}.

\begin{figure}[bth]
	\unitlength1cm
	\begin{picture}(14.0,8.0)   
	\put(0.0,-0.3)               
	{\epsfxsize=15.0cm           
		\epsfbox{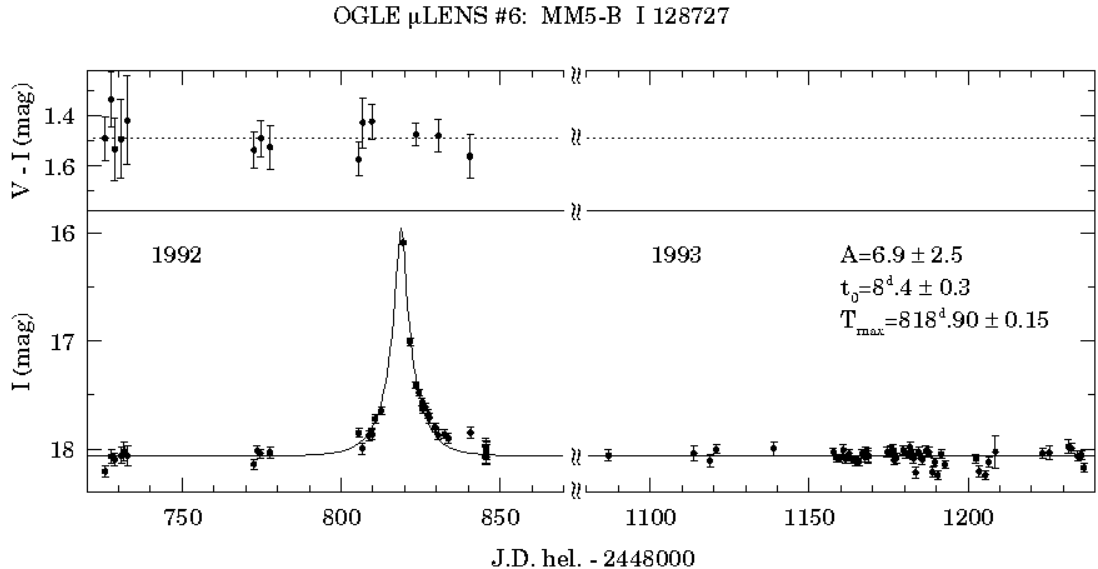}                          
			}   
	\end{picture}
	\caption{\label{fig-bulge-1}
	Observed Lightcurve  of a microlensing event towards 
	the bulge of the galaxy, event OGLE \#6 \cite{ogle94a}:
	The I-band magnitude
	is plotted as a function of time (Julian days).  
	In the top panel
	the constant $V-I$ color of the star is shown. 
	The maximum magnification
	is $\mu = 6.9$ (or 2.1mag), the duration of the event is
	8.4 days. 
	The star has constant brightness in the following year
	(Credits: Andrzej Udalski).
	}
\end{figure}

\begin{figure}[bth]
	\unitlength1cm
	\begin{picture}(14.0,8.5)   
	\put(0.5,-0.1)               
	{\epsfxsize=15.0cm           
		\epsfbox{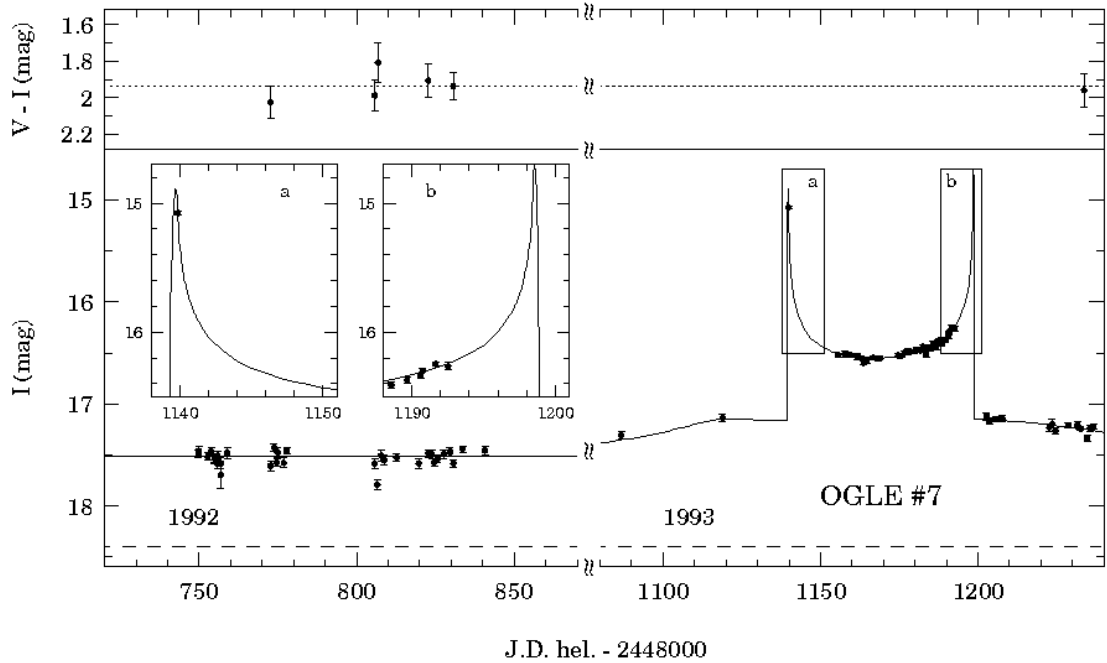}                          
			}   
	\end{picture}
	\caption{\label{fig-bulge-2}
	Lightcurve of a binary microlensing event
	towards the bulge of the galaxy, event OGLE \#7 \cite{ogle94b}:
	The I-band the magnitude
	is plotted over time (Julian days).  In the top panel
	the constant V-I-color of the star is shown. 
	The maximum magnification
	is more than 2.5 mag higher than the unlensed
	brightness. The duration of the event is
	about 80 days. 
	The two insets at the left part show a zoom of the two
	peaks.
	The star had constant brightness in the year preceding  the
	microlensing event (1992). 
	A model  for this event finds a mass ratio of 1.02 between
	the two lensing stars, and a separation of 1.14 Einstein radii
	(Credits: Andrzej Udalski).
}
\end{figure}

The same type of experiment (searching for microlensing events)
is being  performed in the direction of the galactic bulge as well,
the central part of the Milky Way.
By now more than 200 microlensing events have been detected in
this direction (for an example see Figure \ref{fig-bulge-1}). 
Among them are a number of ``binary lens"-events
(which have a very typical signature of at least two caustic crossings,
cf. Figure \ref{fig-bulge-2}).
This is about three times as many microlensing events
as were expected/predicted.   
Several groups try to explain this ``over-abundance" of events
to have a new look at the stellar content and the dynamics
of the bar/bulge of the Galaxy.
The latest published results can be found in \cite{macho97b}.

With these microlensing
experiments gravitational lensing has established
itself as a new tool to study the structure of the Milky Way.
This type of microlensing also holds some promise for the future.
It can be used, e.g. to study the frequency of binary stars. 
One of the most interesting possibilities is to detect planets around
other stars by extending the sensitivity of the binary
lenses to smaller and smaller companion masses  \cite{MP91,Wam97pla}.

%
%

%
%

For a recent comprehensive  presentation of  galactic
microlensing and beyond  see \cite{Pac96}. Various 
aspects of microlensing in the local group are reviewed in detail.
Another review article on the basics and the results of galactic
microlensing  can be found in \cite{RM96}.

%
%
%
%
%
%
%
%
\section{Future Gravitational Lensing}
\label{sec-futu}

Gravitational lensing is an exceptional field in astronomy in 
the sense that its occurence and many of its features  --
e.g.  multiple images, time delays, Einstein rings, 
quasar microlensing, 
galactic microlensing, 
weak lensing --
were {\it predicted} (long) before they were actually observed.
Although ``prediction"  or predictability is considered one 
of the important criteria of modern science, many (astro-)physical
phenomena are too complicated for a minute prediction 
(just think of the weather forecast).
The reason why this worked here is that gravitational
lensing is a simple geometrical concept which
easily allows qualitative estimates and quantitative calculations. 
Extrapolating from these thoughts, 
it should be possible to look forward in time 
once again and predict future applications of gravitational lensing. 

However, at any given time it requires very good intuition, some
courage and maybe even a bit of ingenuity to predict 
qualitatively new phenomena. 
It does not need much of either to envision that
the known lensing phenomena will become better, sharper, more.
My predictions for the next decade in this sense are  humble and modest:

No doubt there will soon be more determinations of accurate
time delays  in multiply-imaged quasar systems.
If the models will get more precise  as well,                the value
of the Hubble constant $H_0$ determined from a number of lens systems
will be accurate to a few percent or better
and will probably turn out to be as reliable as $H_0$ values
obtained
with any other method \cite{WS97}.

The frequencies, image separations, redshift distributions
of multiply-imaged quasars  and their lenses will become a major tool
in differentiating between different cosmological models. 
The Sloan Digital Sky Survey, e.g., will discover a few hundred new
lensed quasars with very well defined selection criteria, ideally
suited for that purpose.
Another angle on the cosmological model and the values of $\Omega$
and $\Lambda$ offer the statistics of arcs. 
The number of high redshift galaxies seen as arcs depends
crucially on the number of rich galaxy clusters at intermediate 
redshifts.  And since different
cosmological models predict very different formation redshifts
for clusters, this  promising road should be followed as 
well \cite{Bar98}.

The new facilities which become available now 
or in the near future in the infrared/sub-mm/mm domain -- like
SCUBA, SIRTF, FIRST, IRAM -- will open a completely new window
in these wavelength ranges, with supposedly most spectacular
results in the arcs and cluster lensing regime.

Quasar microlensing will  provide information on the structure
of the quasars and the intervening clumped matter. With 
the new X-ray telescope AXAF  with its high spatial
resolution it will
become possible to obtain X-ray lightcurves which due to the
presumably smaller emission region will produce dramatic
microlensing events in multiply-imaged quasars. 
Maybe we can ``map" the hot spots of quasars this way.

The largest number of lensing events in the near future will
doubtlessly come from the ``local" microlensing experiments
monitoring galactic bulge stars.
The art of taking spectra of highly magnified stars during
microlensing events (as pioneered by \cite{Len96}) will
open up the fascinating possibility to investigate 
the metallicity of bulge stars in detail
or even resolve the stellar surfaces and
study their center-to-limb variations.
In addition of being an 
excellent tool to study the structure of the Milky Way, galactic
microlensing
will also provide unbiased statistics on the fraction of binary
stars (within certain relative distances). 
Extending the sensitivity to higher mass ratios between the 
binary components will  naturally lead to the detection of
planets around stars (at distances of many kiloparsecs!). 
Microlensing has the advantage compared to all other 
Earth-bound planet search 
techniques that it is able to detect {\it Earth-mass planets}!
It is also imaginable that before too long
such microlensing events could be detected
directly by monitoring astrometrically the position of
the star very accurately \cite{miralda}.

In due course we should also know quantitatively how much 
dark compact objects contribute  to the mass of the halo 
of the Milky Way, and what their mass range is. 
The ``pixel lensing" will probe other lines of sight through 
the Galactic halo by exploring the Andromeda galaxy 
and other nearby galaxies.
This will provide information on the three-dimensional mass
distribution of the halo.

Weak lensing will be used to map not just the outskirts of 
massive galaxy clusters, but also to trace the large scale structure by
its effect on the background population of galaxies. If we find good
ways to 
discriminate between source galaxies at various redshifts, 
this way we can ultimately produce a three-dimensional map of 
the {\it matter} in the universe (rather than a light map)!
This will be an utmost useful thing for the understanding 
of structure formation and evolution; as an aside
we will determine the matter content of the universe  $\Omega$.

Some other possible applications of lensing will be:
The black hole in the Galactic center affects all sources that are 
near or behind the center. Mapping this effect will be a complementary
determination of the black hole mass and will help to study the dynamics
near the black hole.
The redshift of the most distant object will be pushed beyond $z = 6$,
and it is quite likely that it will be magnified by lensing.
The next generation of experiments to map the 
cosmic microwave background 
will be sensititive enough to   detect the gravitational lens signature
of the matter ``in front".

What about the not-so-predictable or 
                    not-so-easily-predictable future of lensing?
Ultimately every object  in the sky is affected by
(ever so slight) lensing effects: this is the not-yet-reached
regime of ultra weak lensing.
I would like to conclude citing two remarks that
Bill Press presented in his lensing outlook 
at the IAU Symposium 173 in Melbourne (1995). 
He mentions that ``gravitational lens effects ... are present
along virtually every line of sight" \cite{Pre95}. In a not
quite so serious extrapolation   Press points out that more and more
astronomers will (have to) deal with lensing in the next decade, so
that lensing will become an ``ubiquitous observational technique" and
hence -- for better or for worse: 
``gravitational lensing
may well disappear as a unique sub-specialty in astronomy".

\noindent

\section*{Acknowledgements}

It is a pleasure to thank 
Wes Colley, 
Frederic Courbin,
Emilio Falco,
Henk Hoekstra,
Neal Jackson,
Tomislav Kundi\'c,
Geraint Lewis, and
Andrzej Udalski
for permission to use their figures.
I would also  like to thank 
Matthias Bartelmann,
Emilio Falco,
Jean-Paul Kneib,
Bohdan Paczy\'nski,
Sjur Refsdal,
Robert Schmidt,
Liliya Williams and
David Woods
for their careful reading of (parts of) the manuscript 
at various stages and their useful comments.
Of particular help were the comments of J\"urgen Ehlers
and an unknown referee
which improved the paper considerably.

\footnotesize

%
%
%
	\def\G     {Gravitational\ }
	\def\Gml   {Gravitational microlensing\ }
	\def\g     {gravitational\ }
	\def\ga    {\gamma}
	\def\gml   {gravitational microlensing\ }
	\def\gl    {gravitational lensing\ }
	\def\gle   {gravitational lens effect\ }
	\def\glt   {gravitational lens theory\ }
	\def\kl#1 {\lbrack #1\rbrack }
	\def\tit#1 {\ }
	\def\etal { {et al.} \ }
	\def\aa#1,#2.{{\sl Astron. Astroph.}, {\bf #1}, #2.}
	\def\actaa#1,#2.{{\sl Acta Astron.}, {\bf #1}, #2.}
	\def\aas#1,#2.{{\sl Astron. Ap. Suppl.}, {\bf #1}, #2.}
	\def\aar#1,#2.{{\sl Astron. Ap. Rev.}, {\bf #1}, #2.}
	\def\aj#1,#2.{{\sl Astron. J.}, {\bf #1}, #2.}
	\def\ag#1,#2.{{\sl Astron. and Geophys.J.}, {\bf #1}, #2.}
	\def\asn#1,#2.{{\sl Astron.~Nachr.}, {\bf #1}, #2.}
	\def\araa#1,#2.{{\sl Ann.~Rev.~Astr.~Ap.}, {\bf #1}, #2.}
	\def\apj#1,#2.{{\sl Astrophys. J.}, {\bf #1}, #2.}
	\def\apjl#1,#2.{{\sl Astrophys. J. Lett.}, {\bf #1}, #2.}
	\def\apjs#1,#2.{{\sl Ap. J. Suppl.}, {\bf #1}, #2.}
	\def\ass#1,#2.{{\sl Astrophys. and Space Sci.}, {\bf #1}, #2.}
	\def\baas#1,#2.{{\sl Bull. Am. Astron. Soc.}, {\bf #1}, #2.}
	\def\coa#1,#2.{{\sl Comments Astrophys. }, {\bf #1}, #2.}
	\def\fcp#1,#2.{{\sl Fundamentals Cosm. Phys.}, {\bf #1}, #2.}
	\def\jaa#1,#2.{{\sl J. Astrophys. Astr.}, {\bf #1}, #2.}
	\def\jgr#1,#2.{{\sl J. Geophys. Res.}, {\bf #1}, #2.}
	\def\jmp#1,#2.{{\sl J. Math. Phys.}, {\bf #1}, #2.}
	\def\livrev#1,#2.{{\sl Liv. Rev. Rel.}, {\bf #1}, #2.}
	\def\mnras#1,#2.{{\sl Mon. Not. R. Astron. Soc.}, {\bf #1}, #2.}
	\def\nat#1,#2.{{\sl Nature}, {\bf #1}, #2.}
	\def\obs#1,#2.{{\sl Observatory}, {\bf #1}, #2.}
	\def\pasp#1,#2.{{\sl Publ.~Astr.~Soc.~Pac.}, {\bf #1}, #2.}
	\def\phyrep#1,#2.{{\sl Phys. Rep.}, {\bf #1}, #2.}
	\def\philmag#1,#2.{{\sl Phil. Mag.}, {\bf #1}, #2.}
	\def\pr#1,#2.{{\sl Phys. Rev.}, {\bf #1}, #2.}
	\def\prb#1,#2.{{\sl Phys. Rev. B}, {\bf #1}, #2.}
	\def\prd#1,#2.{{\sl Phys. Rev. D}, {\bf #1}, #2.}
	\def\prl#1,#2.{{\sl Phys. Rev. Lett.}, {\bf #1}, #2.}
	\def\prsl#1,#2.{{\sl Proc. Roy. Soc. London}, {\bf A#1}, #2.}
	\def\qjras#1,#2.{{\sl Q. Jl R. astr. Soc.}, {\bf #1}, #2.}
	\def\rmp#1,#2.{{\sl Rev. Mod. Phys.}, {\bf #1}, #2.}
	\def\rpp#1,#2.{{\sl Rep. Prog. Phys. }, {\bf #1}, #2.}
	\def\sa#1,#2.{{\sl Sov. Astr.}, {\bf #1}, #2.}
	\def\spd#1,#2.{{\sl Sov. Phys. Doklady.}, {\bf #1}, #2.}
	\def\sci#1,#2.{{\sl Science}, {\bf #1}, #2.}
%
%

\end{document}